\documentclass[11pt]{article}

\usepackage{url}
\usepackage{latexsym}
\usepackage{amssymb}
\usepackage{fullpage}
\newcommand{\SFW}{\mbox{S$^4$W}}
\newcommand{\eobs}{\hspace{\stretch{1}}$\Box$}

\ifx\undefined\psfig\else \fi

\edef\psfigRestoreAt{\catcode`@=\number\catcode`@\relax}
\catcode`\@=11\relax
\newwrite\@unused
\def\typeout#1{{\let\protect\string\immediate\write\@unused{#1}}}
\def\figurepath{./}

\def\@nnil{\@nil}
\def\@empty{}
\def\@psdonoop#1\@@#2#3{}
\def\@psdo#1:=#2\do#3{\edef\@psdotmp{#2}\ifx\@psdotmp\@empty \else
    \expandafter\@psdoloop#2,\@nil,\@nil\@@#1{#3}\fi}
\def\@psdoloop#1,#2,#3\@@#4#5{\def#4{#1}\ifx #4\@nnil \else
       #5\def#4{#2}\ifx #4\@nnil \else#5\@ipsdoloop #3\@@#4{#5}\fi\fi}
\def\@ipsdoloop#1,#2\@@#3#4{\def#3{#1}\ifx #3\@nnil 
       \let\@nextwhile=\@psdonoop \else
      #4\relax\let\@nextwhile=\@ipsdoloop\fi\@nextwhile#2\@@#3{#4}}
\def\@tpsdo#1:=#2\do#3{\xdef\@psdotmp{#2}\ifx\@psdotmp\@empty \else
    \@tpsdoloop#2\@nil\@nil\@@#1{#3}\fi}
\def\@tpsdoloop#1#2\@@#3#4{\def#3{#1}\ifx #3\@nnil 
       \let\@nextwhile=\@psdonoop \else
      #4\relax\let\@nextwhile=\@tpsdoloop\fi\@nextwhile#2\@@#3{#4}}
\newread\ps@stream
\newif\ifnot@eof       
\newif\if@noisy        
\newif\if@atend        
\newif\if@psfile       
{\catcode`\%=12\global\gdef\epsf@start{
\def\epsf@PS{PS}
\def\epsf@getbb#1{%
\openin\ps@stream=#1
\ifeof\ps@stream\typeout{Error, File #1 not found}\else
   {\not@eoftrue \chardef\other=12
    \def\do##1{\catcode`##1=\other}\dospecials \catcode`\ =10
    \loop
       \if@psfile
	  \read\ps@stream to \epsf@fileline
       \else{
	  \obeyspaces
          \read\ps@stream to \epsf@tmp\global\let\epsf@fileline\epsf@tmp}
       \fi
       \ifeof\ps@stream\not@eoffalse\else
       \if@psfile\else
       \expandafter\epsf@test\epsf@fileline:. \\%
       \fi
          \expandafter\epsf@aux\epsf@fileline:. \\%
       \fi
   \ifnot@eof\repeat
   }\closein\ps@stream\fi}%
\long\def\epsf@test#1#2#3:#4\\{\def\epsf@testit{#1#2}
			\ifx\epsf@testit\epsf@start\else
\typeout{Warning! File does not start with `\epsf@start'.  It may not be a PostScript file.}
			\fi
			\@psfiletrue} 
{\catcode`\%=12\global\let\epsf@percent=
\long\def\epsf@aux#1#2:#3\\{\ifx#1\epsf@percent
   \def\epsf@testit{#2}\ifx\epsf@testit\epsf@bblit
	\@atendfalse
        \epsf@atend #3 . \\%
	\if@atend	
	   \if@verbose{
		\typeout{psfig: found `(atend)'; continuing search}
	   }\fi
        \else
        \epsf@grab #3 . . . \\%
        \not@eoffalse
        \global\no@bbfalse
        \fi
   \fi\fi}%
\def\epsf@grab #1 #2 #3 #4 #5\\{%
   \global\def\epsf@llx{#1}\ifx\epsf@llx\empty
      \epsf@grab #2 #3 #4 #5 .\\\else
   \global\def\epsf@lly{#2}%
   \global\def\epsf@urx{#3}\global\def\epsf@ury{#4}\fi}%
\def\epsf@atendlit{(atend)} 
\def\epsf@atend #1 #2 #3\\{%
   \def\epsf@tmp{#1}\ifx\epsf@tmp\empty
      \epsf@atend #2 #3 .\\\else
   \ifx\epsf@tmp\epsf@atendlit\@atendtrue\fi\fi}

\chardef\letter = 11
\chardef\other = 12

\newif \ifdebug 
\newif\ifc@mpute 
\c@mputetrue 

\let\then = \relax
\def\r@dian{pt }
\let\r@dians = \r@dian
\let\dimensionless@nit = \r@dian
\let\dimensionless@nits = \dimensionless@nit
\def\internal@nit{sp }
\let\internal@nits = \internal@nit
\newif\ifstillc@nverging
\def \Mess@ge #1{\ifdebug \then \message {#1} \fi}

{ 
	\catcode `\@ = \letter
	\gdef \nodimen {\expandafter \n@dimen \the \dimen}
	\gdef \term #1 #2 #3%
	       {\edef \t@ {\the #1}
		\edef \t@@ {\expandafter \n@dimen \the #2\r@dian}%
		\t@rm {\t@} {\t@@} {#3}%
	       }
	\gdef \t@rm #1 #2 #3%
	       {{%
		\count 0 = 0
		\dimen 0 = 1 \dimensionless@nit
		\dimen 2 = #2\relax
		\Mess@ge {Calculating term #1 of \nodimen 2}%
		\loop
		\ifnum	\count 0 < #1
		\then	\advance \count 0 by 1
			\Mess@ge {Iteration \the \count 0 \space}%
			\Multiply \dimen 0 by {\dimen 2}%
			\Mess@ge {After multiplication, term = \nodimen 0}%
			\Divide \dimen 0 by {\count 0}%
			\Mess@ge {After division, term = \nodimen 0}%
		\repeat
		\Mess@ge {Final value for term #1 of 
				\nodimen 2 \space is \nodimen 0}%
		\xdef \Term {#3 = \nodimen 0 \r@dians}%
		\aftergroup \Term
	       }}
	\catcode `\p = \other
	\catcode `\t = \other
	\gdef \n@dimen #1pt{#1} 
}

\def \Divide #1by #2{\divide #1 by #2} 

\def \Multiply #1by #2
       {{
	\count 0 = #1\relax
	\count 2 = #2\relax
	\count 4 = 65536
	\Mess@ge {Before scaling, count 0 = \the \count 0 \space and
			count 2 = \the \count 2}%
	\ifnum	\count 0 > 32767 
	\then	\divide \count 0 by 4
		\divide \count 4 by 4
	\else	\ifnum	\count 0 < -32767
		\then	\divide \count 0 by 4
			\divide \count 4 by 4
		\else
		\fi
	\fi
	\ifnum	\count 2 > 32767 
	\then	\divide \count 2 by 4
		\divide \count 4 by 4
	\else	\ifnum	\count 2 < -32767
		\then	\divide \count 2 by 4
			\divide \count 4 by 4
		\else
		\fi
	\fi
	\multiply \count 0 by \count 2
	\divide \count 0 by \count 4
	\xdef \product {#1 = \the \count 0 \internal@nits}%
	\aftergroup \product
       }}

\def\r@duce{\ifdim\dimen0 > 90\r@dian \then   
		\multiply\dimen0 by -1
		\advance\dimen0 by 180\r@dian
		\r@duce
	    \else \ifdim\dimen0 < -90\r@dian \then  
		\advance\dimen0 by 360\r@dian
		\r@duce
		\fi
	    \fi}

\def\Sine#1%
       {{%
	\dimen 0 = #1 \r@dian
	\r@duce
	\ifdim\dimen0 = -90\r@dian \then
	   \dimen4 = -1\r@dian
	   \c@mputefalse
	\fi
	\ifdim\dimen0 = 90\r@dian \then
	   \dimen4 = 1\r@dian
	   \c@mputefalse
	\fi
	\ifdim\dimen0 = 0\r@dian \then
	   \dimen4 = 0\r@dian
	   \c@mputefalse
	\fi
	\ifc@mpute \then
		\divide\dimen0 by 180
		\dimen0=3.141592654\dimen0
		\dimen 2 = 3.1415926535897963\r@dian 
		\divide\dimen 2 by 2 
		\Mess@ge {Sin: calculating Sin of \nodimen 0}%
		\count 0 = 1 
		\dimen 2 = 1 \r@dian 
		\dimen 4 = 0 \r@dian 
		\loop
			\ifnum	\dimen 2 = 0 
			\then	\stillc@nvergingfalse 
			\else	\stillc@nvergingtrue
			\fi
			\ifstillc@nverging 
			\then	\term {\count 0} {\dimen 0} {\dimen 2}%
				\advance \count 0 by 2
				\count 2 = \count 0
				\divide \count 2 by 2
				\ifodd	\count 2 
				\then	\advance \dimen 4 by \dimen 2
				\else	\advance \dimen 4 by -\dimen 2
				\fi
		\repeat
	\fi		
			\xdef \sine {\nodimen 4}%
       }}

\def\Cosine#1{\ifx\sine\UnDefined\edef\Savesine{\relax}\else
		             \edef\Savesine{\sine}\fi
	{\dimen0=#1\r@dian\multiply\dimen0 by -1
	 \advance\dimen0 by 90\r@dian
	 \Sine{\nodimen 0}
	 \xdef\cosine{\sine}
	 \xdef\sine{\Savesine}}}	      

\def\psdraft{
	\def\@psdraft{0}
}
\def\psfull{
	\def\@psdraft{100}
}

\psfull

\newif\if@draftbox
\def\psnodraftbox{
	\@draftboxfalse
}
\@draftboxtrue

\newif\if@prologfile
\newif\if@postlogfile
\def\pssilent{
	\@noisyfalse
}
\def\psnoisy{
	\@noisytrue
}
\psnoisy
\newif\if@bbllx
\newif\if@bblly
\newif\if@bburx
\newif\if@bbury
\newif\if@height
\newif\if@width
\newif\if@rheight
\newif\if@rwidth
\newif\if@angle
\newif\if@clip
\newif\if@verbose
\newif\if@scale
\def\@p@@sclip#1{\@cliptrue}

\def\@p@@sfile#1{\def\@p@sfile{null}%
	        \openin1=#1
		\ifeof1\closein1%
		       \openin1=\figurepath#1
			\ifeof1\typeout{Error, File #1 not found}
			   \if@bbllx\if@bblly\if@bburx\if@bbury
			      \def\@p@sfile{#1}%
			   \fi\fi\fi\fi
			\else\closein1
			    \edef\@p@sfile{\figurepath#1}%
                        \fi%
		 \else\closein1%
		       \def\@p@sfile{#1}%
		 \fi}
\def\@p@@sfigure#1{\def\@p@sfile{null}%
	        \openin1=#1
		\ifeof1\closein1%
		       \openin1=\figurepath#1
			\ifeof1\typeout{Error, File #1 not found}
			   \if@bbllx\if@bblly\if@bburx\if@bbury
			      \def\@p@sfile{#1}%
			   \fi\fi\fi\fi
			\else\closein1
			    \def\@p@sfile{\figurepath#1}%
                        \fi%
		 \else\closein1%
		       \def\@p@sfile{#1}%
		 \fi}

\def\@p@@sbbllx#1{
		\@bbllxtrue
		\dimen100=#1
		\edef\@p@sbbllx{\number\dimen100}
}
\def\@p@@sbblly#1{
		\@bbllytrue
		\dimen100=#1
		\edef\@p@sbblly{\number\dimen100}
}
\def\@p@@sbburx#1{
		\@bburxtrue
		\dimen100=#1
		\edef\@p@sbburx{\number\dimen100}
}
\def\@p@@sbbury#1{
		\@bburytrue
		\dimen100=#1
		\edef\@p@sbbury{\number\dimen100}
}
\def\@p@@sheight#1{
		\@heighttrue
		\dimen100=#1
   		\edef\@p@sheight{\number\dimen100}
}
\def\@p@@swidth#1{
		\@widthtrue
		\dimen100=#1
		\edef\@p@swidth{\number\dimen100}
}
\def\@p@@srheight#1{
		\@rheighttrue
		\dimen100=#1
		\edef\@p@srheight{\number\dimen100}
}
\def\@p@@srwidth#1{
		\@rwidthtrue
		\dimen100=#1
		\edef\@p@srwidth{\number\dimen100}
}
\def\@p@@sangle#1{
		\@angletrue
		\edef\@p@sangle{#1} 
}
\def\@p@@ssilent#1{ 
		\@verbosefalse
}
\def\@p@@sscale#1{
		\def\@p@scale{#1}
		\@scaletrue
}
\def\@p@@sprolog#1{\@prologfiletrue\def\@prologfileval{#1}}
\def\@p@@spostlog#1{\@postlogfiletrue\def\@postlogfileval{#1}}
\def\@cs@name#1{\csname #1\endcsname}
\def\@setparms#1=#2,{\@cs@name{@p@@s#1}{#2}}
\def\ps@init@parms{
		\@bbllxfalse \@bbllyfalse
		\@bburxfalse \@bburyfalse
		\@heightfalse \@widthfalse
		\@rheightfalse \@rwidthfalse
		\@scalefalse
		\def\@p@sbbllx{}\def\@p@sbblly{}
		\def\@p@sbburx{}\def\@p@sbbury{}
		\def\@p@sheight{}\def\@p@swidth{}
		\def\@p@srheight{}\def\@p@srwidth{}
		\def\@p@sangle{0}
		\def\@p@sfile{}
		\def\@p@scost{10}
		\def\@sc{}
		\@prologfilefalse
		\@postlogfilefalse
		\@clipfalse
		\if@noisy
			\@verbosetrue
		\else
			\@verbosefalse
		\fi
}
\def\parse@ps@parms#1{
	 	\@psdo\@psfiga:=#1\do
		   {\expandafter\@setparms\@psfiga,}}
\newif\ifno@bb
\def\bb@missing{
	\if@verbose{
		\typeout{psfig: searching \@p@sfile \space  for bounding box}
	}\fi
	\no@bbtrue
	\epsf@getbb{\@p@sfile}
        \ifno@bb \else \bb@cull\epsf@llx\epsf@lly\epsf@urx\epsf@ury\fi
}	
\def\bb@cull#1#2#3#4{
	\dimen100=#1 bp\edef\@p@sbbllx{\number\dimen100}
	\dimen100=#2 bp\edef\@p@sbblly{\number\dimen100}
	\dimen100=#3 bp\edef\@p@sbburx{\number\dimen100}
	\dimen100=#4 bp\edef\@p@sbbury{\number\dimen100}
	\no@bbfalse
}

\newdimen\p@intvaluex
\newdimen\p@intvaluey
\newdimen\@ffsetvalue
\newdimen\x@ffsetvalue
\newdimen\y@ffsetvalue

\def\compute@offset#1#2{{\dimen0=#1 sp\dimen1=#2 sp
			\advance\dimen1 by -\dimen0
			\dimen1=\sine\dimen1
			\dimen0=\cosine\dimen1
			\ifdim\dimen0<0sp \dimen1=0sp \fi
			\global\@ffsetvalue=\dimen1}}

\def\rotate@#1#2{{\dimen0=#1 sp\dimen1=#2 sp
		  \global\p@intvaluex=\cosine\dimen0
		  \dimen3=\sine\dimen1
		  \global\advance\p@intvaluex by -\dimen3
		  \global\p@intvaluey=\sine\dimen0
		  \dimen3=\cosine\dimen1
		  \global\advance\p@intvaluey by \dimen3
		  }}
\def\compute@bb{
		\no@bbfalse
		\if@bbllx \else \no@bbtrue \fi
		\if@bblly \else \no@bbtrue \fi
		\if@bburx \else \no@bbtrue \fi
		\if@bbury \else \no@bbtrue \fi
		\ifno@bb \bb@missing \fi
		\ifno@bb \typeout{FATAL ERROR: no bb supplied or found}
			\no-bb-error
		\fi
		\if@angle 
			\Sine{\@p@sangle}\Cosine{\@p@sangle}
			\compute@offset{\@p@sbblly}{\@p@sbbury}
			\x@ffsetvalue=\@ffsetvalue
			\compute@offset{\@p@sbburx}{\@p@sbbllx}
			\y@ffsetvalue=\@ffsetvalue

			\rotate@{\@p@sbbllx}{\@p@sbblly}
			\advance\p@intvaluex by -\x@ffsetvalue
			\advance\p@intvaluey by -\y@ffsetvalue
			\edef\@p@sbbllx{\number\p@intvaluex}
			\edef\@p@sbblly{\number\p@intvaluey}

			\rotate@{\@p@sbburx}{\@p@sbbury}
			\advance\p@intvaluex by \x@ffsetvalue
			\advance\p@intvaluey by \y@ffsetvalue
			\edef\@p@sbburx{\number\p@intvaluex}
			\edef\@p@sbbury{\number\p@intvaluey}
			{
			 \count0=\@p@sbbllx \count1=\@p@sbblly
		 	 \count2=\@p@sbburx \count3=\@p@sbbury
			 \dimen0=\@p@sbbllx sp\dimen1=\@p@sbblly sp
		 	 \dimen2=\@p@sbburx sp\dimen3=\@p@sbbury sp
			 \dimen203=\dimen2 \advance\dimen203 by -\dimen0
			 \dimen204=\dimen3 \advance\dimen204 by -\dimen1
			 \ifdim\dimen203<0sp 
			      \count203=\count2 \count2=\count0 
			      \count0=\count203 
			      \global\edef\@p@sbbllx{\number\count0}
			      \global\edef\@p@sbburx{\number\count2}
			 \fi
			 \ifdim\dimen204<0sp 
			       \count204=\count3
			       \count3=\count1
			       \count1=\count204
			       \global\edef\@p@sbblly{\number\count1}
			       \global\edef\@p@sbbury{\number\count3}
			 \fi
			}
		\fi
		\count203=\@p@sbburx
		\count204=\@p@sbbury
		\advance\count203 by -\@p@sbbllx
		\advance\count204 by -\@p@sbblly
		\edef\@bbw{\number\count203}
		\edef\@bbh{\number\count204}
}
\def\in@hundreds#1#2#3{\count240=#2 \count241=#3
		     \count100=\count240	
		     \divide\count100 by \count241
		     \count101=\count100
		     \multiply\count101 by \count241
		     \advance\count240 by -\count101
		     \multiply\count240 by 10
		     \count101=\count240	
		     \divide\count101 by \count241
		     \count102=\count101
		     \multiply\count102 by \count241
		     \advance\count240 by -\count102
		     \multiply\count240 by 10
		     \count102=\count240	
		     \divide\count102 by \count241
		     \count200=#1\count205=0
		     \count201=\count200
			\multiply\count201 by \count100
		 	\advance\count205 by \count201
		     \count201=\count200
			\divide\count201 by 10
			\multiply\count201 by \count101
			\advance\count205 by \count201
		     \count201=\count200
			\divide\count201 by 100
			\multiply\count201 by \count102
			\advance\count205 by \count201
		     \edef\@result{\number\count205}
}
\def\@ScaleInHundreds#1{
		\in@hundreds{#1}{\@p@scale}{100}
		\edef#1{\@result}
}
\def\compute@wfromh{
		\in@hundreds{\@p@sheight}{\@bbw}{\@bbh}
		\edef\@p@swidth{\@result}
}
\def\compute@hfromw{
		\in@hundreds{\@p@swidth}{\@bbh}{\@bbw}
		\edef\@p@sheight{\@result}
}
\def\compute@handw{
		\if@height 
			\if@width
			\else
				\compute@wfromh
			\fi
		\else 
			\if@width
				\compute@hfromw
			\else
				\edef\@p@sheight{\@bbh}
				\edef\@p@swidth{\@bbw}
			\fi
		\fi
}
\def\compute@resv{
		\if@rheight \else \edef\@p@srheight{\@p@sheight} \fi
		\if@rwidth \else \edef\@p@srwidth{\@p@swidth} \fi
}
\def\compute@sizes{
	\compute@bb
	\compute@handw
	\compute@resv
}
\def\psfig#1{\vbox {
	\ps@init@parms
	\parse@ps@parms{#1}
	\compute@sizes
	\if@scale
                \if@verbose
                        \typeout{psfig: scaling by \@p@scale}
                \fi
                \@ScaleInHundreds{\@p@swidth}
                \@ScaleInHundreds{\@p@sheight}
                \@ScaleInHundreds{\@p@srwidth}
                \@ScaleInHundreds{\@p@srheight}
        \fi
	\ifnum\@p@scost<\@psdraft{
		\if@verbose{
			\typeout{psfig: including \@p@sfile \space }
		}\fi
		\special{ps::[begin] 	\@p@swidth \space \@p@sheight \space
				\@p@sbbllx \space \@p@sbblly \space
				\@p@sbburx \space \@p@sbbury \space
				startTexFig \space }
		\if@angle
			\special {ps:: \@p@sangle \space rotate \space} 
		\fi
		\if@clip{
			\if@verbose{
				\typeout{(clip)}
			}\fi
			\special{ps:: doclip \space }
		}\fi
		\if@prologfile
		    \special{ps: plotfile \@prologfileval \space } \fi
		\special{ps: plotfile \@p@sfile \space }
		\if@postlogfile
		    \special{ps: plotfile \@postlogfileval \space } \fi
		\special{ps::[end] endTexFig \space }
		\vbox to \@p@srheight true sp{
			\hbox to \@p@srwidth true sp{
				\hss
			}
		\vss
		}
	}\else{
		\if@draftbox{		
			\hbox{\fbox{\vbox to \@p@srheight true sp{
			\vss
			\hbox to \@p@srwidth true sp{ \hss \@p@sfile \hss }
			\vss
			}}}
		}\else{
			\vbox to \@p@srheight true sp{
			\vss
			\hbox to \@p@srwidth true sp{\hss}
			\vss
			}
		}\fi

	}\fi
}}
\def\psglobal{\typeout{psfig: PSGLOBAL is OBSOLETE; use psprint -m instead}}
\psfigRestoreAt

\newif\ifpdf
\ifx\pdfoutput\undefined
  \pdffalse
\else
  \pdfoutput=1
  \pdftrue
\fi

\ifpdf
  \usepackage[pdftex]{graphicx}
  \usepackage[pdftex]{color}
  \DeclareGraphicsExtensions{.pdf,.png,.jpg}
\else
  \usepackage[dvips]{graphicx}
  \usepackage[dvips]{color}
  \DeclareGraphicsExtensions{.eps,.epsi,.ps}
\fi

\usepackage{times}

\def\midv{\mathop{\,|\,}}

\long\def\cbk#1{{\color{red}[CBK: #1]}}
\newlength\colwidth \setlength\colwidth{3.25in}
\begin{document}
\thispagestyle{empty}
\title{BSML: A Binding Schema Markup Language for\\
Data Interchange in Problem Solving Environments\footnote{The work presented in this 
paper is supported in part by National Science Foundation 
grants EIA-9974956, EIA-9984317, and
EIA-0103660.}}
\author{\large Alex Verstak$^*$, Naren Ramakrishnan$^*$, Layne T. Watson$^*$, Jian He$^*$, Clifford A. Shaffer$^*$,\\
\large Kyung Kyoon Bae$^{\dagger}$, Jing Jiang$^{\dagger}$, William H. Tranter$^{\dagger}$, and Theodore S. Rappaport$^{\dagger}$\\
\large $^*$Department of Computer Science\\
\large $^{\dagger}$Bradley Department of Electrical and Computer Engineering\\
\large Virginia Polytechnic Institute and State University\\
\large Blacksburg, Virginia 24061\\
\large Contact: \url{naren@cs.vt.edu}}
\date{}
\maketitle
\vspace{-0.3in}
\begin{abstract}
\noindent
We describe a binding schema markup language (BSML) for describing data
interchange between scientific codes. Such a facility is an important
constituent of scientific problem solving environments (PSEs).
BSML is designed to integrate with a PSE or application
composition system that views model specification and execution 
as a problem of managing semistructured data.  The data interchange problem is addressed by
three techniques for processing semistructured data: validation, binding,
and conversion.  We present BSML and describe its application 
to a PSE for wireless communications system design.
\end{abstract}

\section{Introduction} \label{sec:intro}

Problem solving environments (PSEs) are high-level software systems
for doing computational science.  A simple example of a PSE is the
Web PELLPACK system~\cite{pellpack-cse} that addresses the domain
of partial differential equations (PDEs).  Web PELLPACK allows the
scientist to access the system through a Web browser, define PDE
problems, choose and configure solution strategies, manage appropriate
hardware resources (for solving the PDE), and visualize and analyze
the results.  The scientist thus communicates with the PSE in the
vernacular of the problem, `not in the language of a particular
operating system, programming language, or network protocol'~\cite{jrrpse}.
It is 10 years since the goal of creating PSEs was articulated by an
NSF workshop (see~\cite{jrrpse} for findings and recommendations).
From providing high-level programming interfaces for widely used
software libraries~\cite{RB96}, PSEs have now expanded to diverse
application domains such as wood-based composites design~\cite{wbcsim},
aircraft design~\cite{vizcraft}, gas turbine dynamics
simulation~\cite{cacmpaper}, and microarray bioinformatics~\cite{expresso-hpc}.

The basic functionalities expected of a PSE include supporting the
specification, monitoring, and coordination of extended problem solving
tasks. Many PSE system designs employ the {\it compositional modeling}
paradigm, where the scientist describes data-flow relationships between
codes in terms of a graphical network and the PSE manages the details
of composing the application represented by the network. Compositional
modeling is not restricted to such model specification and execution
but can also be used as an aid in performance modeling of scientific
codes~\cite{poems-tse} (model analysis).

We view model specification and execution as a data management problem
and describe how a semistructured data model can be used to address
data interchange problems in a PSE.  Section~\ref{sec:example} presents
a motivating PSE scenario that will help articulate needs from a data
management perspective.  Section~\ref{sec:reqs} elaborates on these
ideas and briefly reviews pertinent related work. In particular, it
identifies three basic levels of functionality---validation, binding,
and conversion---at which data interchange in application composition
can be studied.  Sections~\ref{sec:validation}, \ref{sec:binding},
and~\ref{sec:conversion} describe our specific contributions along
these dimensions, in the form of a binding schema markup language
(BSML).  Section~\ref{sec:integration} outlines how these ideas can be
integrated within an existing PSE system design. A concluding
discussion is provided in Section~\ref{sec:discussion}.  Aspects of the
scenario described next will be used throughout this paper as running
examples.

\subsection{Motivating Example} \label{sec:example}

\SFW{} (Site-Specific System Simulator for Wireless system design) is a
PSE being developed at Virginia Tech.  \SFW{} provides deterministic
electromagnetic propagation and stochastic wireless system models for
predicting the performance of wireless systems in specific environments,
such as office buildings.  \SFW{} is also designed to support the
inclusion of new models into the system, visualization of results
produced by the models, integration of optimization loops around the
models, validation of models by comparison with field measurements, and
management of the results produced by a large series of experiments.
\SFW{} permits a variety of usage scenarios.  We will describe one
scenario in detail.

A wireless design engineer uses \SFW{} to study transmitter placement
in an indoor environment located on the fourth floor of Durham Hall at
Virginia Tech. The engineering goal is to achieve a certain performance
objective within the given cost constraints.  For a narrowband system,
power levels at the receiver locations are good indicators of system
performance.  Therefore, minimizing the (spatial) average shortfall of
received power with respect to some power threshold is a meaningful and
well defined objective.  The major cost constraints are the number of
transmitters and their powers.  Different transmitter locations and
powers yield different levels of coverage.  The situation is more
complicated in a wideband system, but roughly the same process
applies.  A wideband system includes extra hardware not present in a
narrowband system and the performance objective is formulated in terms
of the bit error rate (BER), not just the power level.

\begin{figure}[tbp]
\begin{center}
\includegraphics[width=5.64in,height=2.139in]{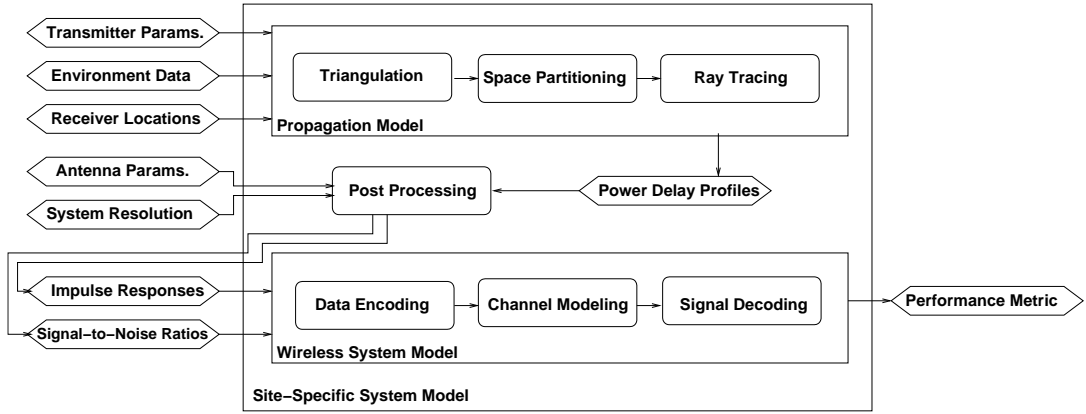}
\end{center}
\caption{A site-specific system model in \SFW.  The system model
consists of a propagation model, an antenna model (post processing),
and a wireless system model.}
\label{fig:loop1}
\end{figure}

The first step in this scenario is to construct a model of signal
propagation through the wireless communications channel.  \SFW{}
provides ray tracing as the primary mechanism to model site-specific
propagation effects such as transmission (penetration), reflection, and
diffraction.  The second step is to take into account antenna
parameters and system resolution.  These two steps are often sufficient
to model the performance of a narrowband system.  If a wideband system
is being considered, the third step is to configure the specific
wireless system.  Parameters such as the number of fingers of the rake
receiver and forward error correction codes are considered at this
step.  \SFW{} provides a Monte-Carlo simulation of a WCDMA (wideband
code division multiple access) family of wireless systems.  In either
case, the engineer configures a graph of computational components as
shown in Fig.~\ref{fig:loop1}.  The ovals correspond to computational
components drawn from a mix of languages and environments.  Hexagons
enclose input and output data.  Aggregation is used to simplify the
interfaces of the components to each other and to the optimizer.  In
Fig.~\ref{fig:loop1}, rectangles represent aggregation.  The
propagation model is a component that consists of three connected
subcomponents: triangulation, space partitioning, and ray tracing.
Similarly, the wireless system model consists of (roughly) three
components: data encoding, channel modeling, and signal decoding.  All
three steps are further aggregated into a complete site-specific system
model.  This model is then used in an optimization loop.  The
optimizer changes transmitter parameters (all other parameters remain
fixed) and receives feedback on system performance.

For a given environment definition in AutoCAD, the triangulation and
space partitioning components are used to reduce the number of
geometric intersection tests that will be performed by the ray tracer.
Several iterations over space partitioning are necessary to achieve
acceptable software performance.  However, once the objective (an
average of ten triangles per voxel) is met, the space partitioning can
be reused in all future experiments with this environment.  The
engineer then configures the ray tracer to only capture reflection and
transmission (penetration) effects.  Although diffraction and
scattering are important in indoor propagation~\cite{ted-prop}, these
phenomena are computationally expensive to model in an optimization
loop.  The triangulation and space partitioning codes are meant for
serial execution, whereas the ray tracer and the Monte Carlo wireless
system models run on a 200 node Beowulf cluster of workstations.
Post processing is available in both serial and parallel versions.  The
ray tracer and the post processor are written in C, whereas the WCDMA
simulation is available in Matlab and Fortran~95 versions.

\begin{figure}[tbp]
\begin{center}
\includegraphics[width=3.2in,height=1in]{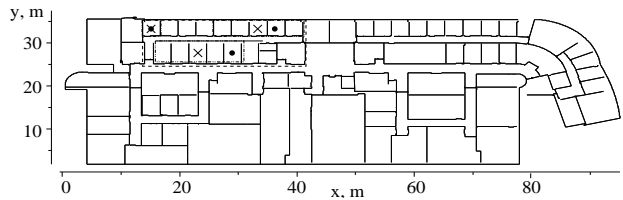}
\end{center}
\caption{Optimizing placement of three
transmitters to cover eighteen rooms and a corridor bounded by the box
in the upper left corner.  The bounds for the placement of three
transmitters are drawn with dotted lines.  The initial transmitter
positions are marked with crosses.  The optimum coverage transmitter
positions are marked with dots.} \label{fig:loop2}
\end{figure}

A series of experiments is performed for various choices of antenna
patterns, path loss parameters (influenced by material properties), and
WCDMA system parameters.  The predicted power delay profiles (PDPs) are then
compared with the measurements from a channel sounder and the predicted
bit error rates are compared with the published data.  The parameters
of the propagation model are calibrated for various locations.  The
validated propagation and wireless system models are finally enclosed
in an optimization loop to determine the locations of transmitters that
will provide adequate performance for a region of interest.  The
optimizer, written in Fortran~95, uses the DIviding RECTangles (DIRECT)
algorithm of Jones et al.~\cite{jones-direct}.  The parameters to the
optimization problem and the optimal transmitter placement are depicted
in Fig.~\ref{fig:loop2}.  The optimizer decided to move the
transmitter in the upper right corner one room to the right of its
initial position and the transmitter in the lower left corner two rooms
to the right of its initial position.

What requirements can we abstract from this scenario and how can they
be flexibly supported by a data model?  We first observe the diversity
in the computational environment.  Component codes are written in
different languages and some of them are meant for parallel execution.
In a research project such as \SFW, many components are under active
development, so their I/O specifications change over time.  Second, the
interconnection among components is also flexible.  Optimizing for
power coverage and optimizing for bit error rate, while having similar
motivations, require different topologies of computational components.
Third, since different groups of researchers are involved in the
project, there exists significant cognitive discordance among
vocabularies, data formats, components, and even methodologies.  For
example, ray tracing models represent powers in a power delay profile
in dBm (log scale).  However, WCDMA models work with a normalized
linear scale impulse response and an aggregate called the
`energy-to-noise ratio.'  Also, there is more than one way of
calculating the energy-to-noise ratio.  Since antennas generate noise
that depends on their parameters, detailed antenna descriptions are
necessary to calculate this ratio.  However, researchers who are not
concerned with antenna design seldom model the system at this level of
detail.  The typical practice is to use a fixed noise level in the
calculations.  Simulations of wireless systems abound in such
approximations, ad hoc conversions, and simplifying assumptions.

\section{PSE Requirements for Data Interchange} \label{sec:reqs}

Culling from the above scenario, we arrive at a more formal list of
data interchange requirements for application composition in a PSE.
The PSE must support:

\begin{enumerate}
\baselineskip=9.9pt
\itemsep=0.02in
\parsep=-0.5in
\item components in multiple languages (C, FORTRAN, Matlab, SQL);
\item changes in component interfaces;
\item changes in interconnections among components;
\item automatic unit conversion in data-flows;
\item user-defined conversion filters;
\item composition of components with slightly different interfaces; and
\item stream processing.
\end{enumerate}

The reader might be surprised that SQL is listed alongside FORTRAN, but
both languages are used in \SFW.  Experiment simulations are written in
procedural languages, while experiment data is stored in a relational
database.  Thus, developing a system that integrates with the PSE
environment requires more than the ability to link scientific computing
languages.  It involves overcoming the impedance mismatch between
languages developed for fundamentally different purposes.

The last requirement above is related to composability---the ability to
create arbitrary component topologies.  As data interchange is pushed
deeper into the computation, the unit of data granularity needs to
become correspondingly smaller.  The optimization loop is a good
example of fine data granularity.  We cannot accumulate all transmitter
parameters over all iterations and later convert them to the format
required by the simulation inside the loop, because transmitter
parameters generated by the optimizer depend on the feedback computed
by the simulation.  Each block of transmitters must be processed as
soon as it is available.  Likewise, each value of the objective
function must be made available to the optimizer before it can produce
the next block of transmitters.  Usability dictates a similar
requirement.  Since some models are computationally expensive (e.g.,
those meant for parallel execution), incremental feedback should be
provided to the user as early as possible.  The stream processing
requirement improves composability and usability, but limits
conversions to being local.  Global conversions (e.g.,
XSLT~\cite{xslt}) cannot be performed because they assume that all the
data is available at once.

While the requirements point to a semistructured data model, no
currently available data management system supports all forms of PSE
functionality. This paper presents the prototype of such a system in
the form of a markup language.  Observe that all of the above
requirements are summarized by three standard techniques for working
with semistructured data---validation, binding, and conversion.
\emph{Validation} establishes data conformance to a given schema.  It
is a prerequisite to most of the requirements.  \emph{Binding} refers
to integrating semistructured data with languages that were designed
for different purposes (requirement~1).  \emph{Conversion}
(transformation) takes care of requirements~2--6.  Given two slightly
different schemas, it is possible to generate an \emph{edit
script}~\cite{chawathe97} that converts data instances from one schema
to another.  Requirement~7 dictates that all such conversions must be
local.

\subsection{Related Work}

While research in PSEs covers a broad territory, the use of
semistructured data representations in computational science is not
established beyond a few projects.  Therefore, we only survey standard
XML technologies and PSE-like systems that make (some) use of
semistructured data. It would be unfair to review some of these systems
against PSE data interchange requirements. Instead, our evaluation is
based on how well these systems support validation, binding,
conversion, and stream processing.

Specific XML technologies for document processing are easy to classify
in terms of our framework. \emph{Schema languages} (e.g.,
RELAX~NG~\cite{relax-ng}) deal with validation and, possibly, binding.
\emph{Transformation languages} (e.g., XSLT~\cite{xslt}) deal with
conversion.  Several properties of these technologies hinder their
direct applicability to a PSE setting.  First and foremost, these
technologies do not work with streams of data.  Sophisticated schema
constraints and complex transformations can require buffering the whole
document before producing any output.  Second, transformation languages
are simply vehicles for applying edit scripts.  They cannot be used to
create edit scripts.  Since our conversions are local, edit script
application is trivial, but edit script creation is not.

Four major flavors of PSE-like projects that use semistructured data
representations can be identified:

\begin{enumerate}
\baselineskip=9.9pt
\itemsep=0.02in
\parsep=-0.5in
\item component metadata projects;
\item workflow projects;
\item scientific data interchange projects; and
\item scientific data management projects.
\end{enumerate}

Projects in the first category use XML to store IDL-like (interface
definition language) component descriptions and miscellaneous component
execution parameters.  An example of such a project is
CCAT~\cite{ccat00}, which is a distributed object oriented system.
CCAT also uses XML for message transport between components, so we say
that it provides an OO binding.  The second category of projects
augments component metadata with workflow specifications.  For example,
GALE~\cite{gale} is a workflow specification language for executing
simulations on distributed systems.  Unlike CCAT, GALE provides XML
specifications for some common types of experiments, such as parameter
sweeps (CCAT uses a scripting language for workflow specification).
However, GALE does not use XML for component data.  Both the component
metadata and workflow projects use XML to encode data that is not
semistructured.  Their use of XML is not dictated by the need for
automatic conversion.  Neither generic binding mechanisms nor
conversion are provided by these projects.

The latter two groups of projects use XML for application data, not
component metadata.  Representatives of the scientific data interchange
group develop flexible all-encompassing schemas for specific
application domains.  For example, CACTUS~\cite{cactus-dd} deals with
spatial grid data.  CACTUS's schema is complex enough to be considered
semistructured and this project recognizes the need for conversion
filters.  However, it does not provide multiple language support and,
more importantly, does not accommodate changes in the schema.
CACTUS's conversion filters aim at code reuse, not change management.
This project has OO binding and manual conversion (the sequence of
conversions is not determined automatically).  Complexity of the data
format precludes stream processing.

Perhaps the most relevant group of projects for our purposes involves
the scientific data management community.  Especially interesting are
the projects in rapidly evolving domains, such as bioinformatics.
DataFoundry~\cite{datafoundry-cd,datafoundry-med} provides a unifying
database interface to diverse bioinformatics sources.  Both the data
and the schema of these sources evolve quickly, so DataFoundry has to
deal with change management---by far more complex change management
than the kind we consider here.  However, DataFoundry only provides
\emph{mediators} for database access.  It does not integrate with
simulation execution.  This system takes full advantage of conversion,
but provides only an SQL binding.  Introducing bindings for procedural
languages would involve significant changes to DataFoundry.

\begin{table}[tbp]
\begin{center}
\begin{tabular}{l c c c c c c}
\hline
           & CCAT    & GALE    & CACTUS  & DataFoundry & RELAX NG & XSLT \\
\hline
Validation & $\surd$ &         & $\surd$ & $\surd$     & $\surd$  & \\
Binding    & OO      &         & OO      & SQL         & OO       & \\
Conversion &         &         & manual  & $\surd$     &          & manual \\
Stream Processing & $\surd$&          &             &          & \\
\hline
\end{tabular}
\end{center}
\caption{A survey of PSE-like systems and XML technologies.  The
binding row shows that most systems support only one paradigm.  Only
DataFoundry fully supports conversion.  Other systems either provide a
library of conversion primitives and leave their composition up to the
user (CACTUS) or do not recognize the need for conversion at all
(CCAT).  No system or technology fully supports validation, binding,
and conversion.  Most systems and technologies cannot dynamically
process streams of data.}
\label{tab:comparison}
\end{table}

Table~\ref{tab:comparison} summarizes related work.  It turns out that
no known PSE-like system takes full advantage of both binding and
conversion.  XML technologies for validation and binding are well
established, but XML transformation technologies do not support
PSE-style conversion.  Very few systems can integrate with a PSE
execution environment because most of them do not meet the stream
processing requirement.  This paper develops a system that satisfies
all of our data interchange requirements.  The next three sections
describe our handling of validation, binding, and conversion.  System
integration is outlined in Section~\ref{sec:integration}.

\section{Validation} \label{sec:validation}

Validation establishes conformance of a data instance to a given
schema.  It is a prerequisite to binding and conversion.  (This
definition of validation is a small part of the process of validation
in a PSE, which is concerned with the larger issue of a model being
appropriate to solve a given problem; but, it suffices for the purpose
of this paper.)  The schemas for PSE data are easy to obtain since
computational science traditionally uses rigid data structures, not
loosely formatted documents.  Describing the data structures in terms
of schemas has several benefits.  First, language-neutral schemas allow
for interoperability between different languages (see requirement~1 in
the previous section).  Second, schemas facilitate database storage and
retrieval.  Third, appropriate schemas help assign interpretations to
various data fields.  It is such interpretation that makes automatic
conversion possible (requirements~2--6).

What kind of validation is appropriate for PSE data?  Requirement~7
calls for the most expressive schema language that can be parsed by a
stream parser.  In other words, we are looking for a schema language
that can be defined in terms of an LL(1) grammar~\cite{dragon-book}.
(The LR family of grammars is more expressive, but LR parsers do not
follow stream semantics.) Therefore, a predictive parser generated for
a given schema can validate a data instance.  This section describes a
schema language (BSML) appropriate for a PSE and the steps for building
a parser generator for this language.  We present an example, an informal
overview of BSML features, and a formal definition for a large subset of
BSML in terms of a context-free grammar.  Further, predictive parser
generation is outlined and grammar transformations specific to BSML are
described in detail.  Finally, we show that BSML is strictly less
expressive than LL(1) grammars.

Let us start with an example.  Figures~\ref{fig:octree-schema}
and~\ref{fig:octree-schema-nxml} depict a (simplified) schema for an
octree environment decomposition.  (Fig.~\ref{fig:octree-schema}
describes it in XML notation while Fig.~\ref{fig:octree-schema-nxml}
uses a non-XML format that will be useful for describing some
functionalities of BSML).  This is the most complex schema in \SFW, not
counting the schema for the schema language itself.  An octree consists
of internal and leaf nodes that delimit groups of triangles.  Recall
from Section~\ref{sec:example} that this grouping is used to limit the
intersection tests in ray tracing.  The nested structure of an octree
maps nicely into an XML tree.  Since many components work with lists of
triangles, there is a separate schema for a list of triangles.  As the
example shows, the features of BSML closely resemble those of other
schema languages, such as RELAX~NG.  The only noticeable difference is
the presence of units in the definitions of primitive types.  Units
will be useful for certain types of conversions.
Figure~\ref{fig:octree-grammar} shows an LL(1) grammar generated from
the octree schema.  This grammar is then annotated with binding code
and used to generate a parser for octree data.  The parser can be
linked with a parallel ray tracer written in C.

\begin{figure}[tbp]
{\small\begin{verbatim}
<type id='distance' base='double' number='true' finite='true'/>
<type id='coordinate' base='double' number='true' finite='true'/>

<schema id='triangles'>
  <repetition>
    <element name='tr'>
      <repetition min='3' max='3'>
        <element name='v'>
          <attribute name='x' type='coordinate' units='m'/>
          <attribute name='y' type='coordinate' units='m'/>
          <attribute name='z' type='coordinate' units='m'/>
        </element>
      </repetition>
    </element>
  </repetition>
</schema>

<schema id='octree'>
  <element name='octree'>
    <element name='oi' id='oi'>
      <attribute name='x' type='coordinate' units='m'/>
      <attribute name='y' type='coordinate' units='m'/>
      <attribute name='z' type='coordinate' units='m'/>
      <attribute name='dx' type='distance' units='m'/>
      <attribute name='dy' type='distance' units='m'/>
      <attribute name='dz' type='distance' units='m'/>
      <ref id='triangles'/>
      <repetition>
        <selection>
          <ref id='oi'/>
          <element name='ol'>
            <attribute name='x' type='coordinate' units='m'/>
            <attribute name='y' type='coordinate' units='m'/>
            <attribute name='z' type='coordinate' units='m'/>
            <attribute name='dx' type='distance' units='m'/>
            <attribute name='dy' type='distance' units='m'/>
            <attribute name='dz' type='distance' units='m'/>
            <ref id='triangles'/>
          </element>
        </selection>
      </repetition>
    </element>
  </element>
</schema>
\end{verbatim}}
\caption{BSML schemas for an octree decomposition of an environment, in
XML notation.  `tr' stands for a triangle, `v' stands for a vertex,
`oi' stands for an internal node, and `ol' stands for a leaf.}
\label{fig:octree-schema}
\end{figure}

\begin{figure}[tbp]
{\small\begin{verbatim}
type(distance, double, $, $, true, true, $)
type(coordinate, double, $, $, true, true, $)

schema(triangles,
  repetition($, $, $, $,
    element($, $, tr,
      repetition($, $, 3, 3,
        element($, $, v,
          attribute($, x, data(coordinate,$,$,$,$,m)),
          attribute($, y, data(coordinate,$,$,$,$,m)),
          attribute($, z, data(coordinate,$,$,$,$,m))
        )
      )
    )
  )
)

schema(octree,
  element($, $, octree,
    element(oi, $, oi,
      attribute($, x, data(coordinate,$,$,$,$,m)),
      attribute($, y, data(coordinate,$,$,$,$,m)),
      attribute($, z, data(coordinate,$,$,$,$,m)),
      attribute($, dx, data(coordinate,$,$,$,$,m)),
      attribute($, dy, data(coordinate,$,$,$,$,m)),
      attribute($, dz, data(coordinate,$,$,$,$,m)),
      ref(triangles),
      repetition($, $, $, $,
        selection($, $,
          ref(oi),
          element($, $, ol,
            attribute($, x, data(coordinate,$,$,$,$,m)),
            attribute($, y, data(coordinate,$,$,$,$,m)),
            attribute($, z, data(coordinate,$,$,$,$,m)),
            attribute($, dx, data(coordinate,$,$,$,$,m)),
            attribute($, dy, data(coordinate,$,$,$,$,m)),
            attribute($, dz, data(coordinate,$,$,$,$,m)),
            ref(triangles)
          )
        )
      )
    )
  )
)
\end{verbatim}}
\caption{BSML schemas from Figure~\ref{fig:octree-schema} in a non-XML
notation.  {\tt \$} stands for a missing value, i.e., a suitable
default value is supplied by BSML software.}
\label{fig:octree-schema-nxml}
\end{figure}

\begin{figure}[tbp]
\begin{center}
\fbox{
$\begin{array}{l c l}
S & \rightarrow & s(octree), s(oi), T, C, e(oi), e(octree) \\
T & \rightarrow & \epsilon \\
T & \rightarrow & \{B_t\}, s(tr), \{B_v\}, s(v), e(v), \{A_v\}, V, \{E_v\}, e(tr), \{A_t\}, T', \{E_t\} \\
T' & \rightarrow & \epsilon \\
T' & \rightarrow & s(tr), \{B_v\}, s(v), e(v), \{A_v\}, V, \{E_v\}, e(tr), \{A_t\}, T' \\
V & \rightarrow & \epsilon \\
V & \rightarrow & s(v), e(v), \{A_v\}, V \\
C & \rightarrow & \epsilon \\
C & \rightarrow & \{B_i\}, C', \{A_i\}, C'', \{E_i\} \\
C' & \rightarrow & s(oi), T, C, e(oi) \\
C' & \rightarrow & s(ol), T, e(ol) \\
C'' & \rightarrow & \epsilon \\
C'' & \rightarrow & I \\
I & \rightarrow & s(oi), T, I' \\
I & \rightarrow & s(ol), T, e(ol), \{A_i\}, C'' \\
I' & \rightarrow & \{B_i\}, C', \{A_i\}, C'', \{E_i\}, e(oi), \{A_i\}, C'' \\
I' & \rightarrow & e(oi), \{A_i\}, C'' \\
\end{array}$}
\end{center}
\caption{LL(1) grammar corresponding to the octree schemas in
Figures~\ref{fig:octree-schema} and~\ref{fig:octree-schema-nxml}.
Attributes are omitted for simplicity.  Patterns of the form $\{c\}$
will be explained in the next section (they are related to
repetitions).  Non-terminals $T$, $T'$, and $V$ are related to
triangles; others are related to octree decomposition of a set of
triangles.} \label{fig:octree-grammar}
\end{figure}

The DTD for the current version of BSML is given in
Appendix~\ref{app:dtd}.  The schema language describes primitive types
and schemas.  There are four base primitive types: integer, string,
(IEEE) double, and boolean.  Users can derive their own primitive types
by range restriction.  User-derived types usually have domain-specific
flavor, such as coordinates and distances in the example above.  We do
not support more complicated primitive types, such as dates and lists,
because each PSE component treats them differently.  Schemas consist of
four building blocks: elements, sequences, selections, and
repetitions.  Strictly speaking, repetitions can be expressed as
selections and sequences, but they are so common that they deserve
special treatment.  Derivation of schemas by restriction is not
supported, but derivation by extension can be implemented via
inter-schema references.  Mixed content is not supported because it is
only used for documentation.  Instead, BSML supports a wildcard content
type.  The contents of this type matches anything and is delivered to
the component as a DOM tree~\cite{dom}.  We do not support referential
integrity constraints because they can delay binding and thus break
requirement~7.  There is no explicit construct for interleaves.  In
some ways, interleaves are handled by the conversion algorithm.  In
other words, BSML is a simple schema language that incorporates most
common features that are useful in a PSE.

Parser generation for a BSML schema follows the standard steps from
compiler textbooks~\cite{dragon-book}:

\begin{enumerate}
\baselineskip=9.9pt
\itemsep=0.02in
\parsep=-0.5in
\item convert the schema to an LL(1) grammar,
\item eliminate empty productions and self-derivations,
\item eliminate left recursion,
\item perform left factoring,
\item perform miscellaneous cleanup (described in detail below),
\item compute a predictive parsing table, and
\item generate parsing code from the table.
\end{enumerate}

The only steps specific to this schema language are generating an LL(1)
grammar (step~1) and miscellaneous cleanup (step~5). Since grammars
have been in use for a long time, it is pertinent to define BSML
semantics in terms of how the schemas are converted to grammars.  The
terminals are defined by SAX events~\cite{sax}.  The start of element
and end of element events are denoted $s(name)$ and $e(name)$,
respectively, where $name$ is element name.  We omit the attributes for
simplicity, but BSML supports them in an obvious way.  Further, we
assume that the SAX parser inlines external entity references.
Character data is accumulated until the next start of element or end of
element event and delivered as a $d(base,min,max,number,finite,units)$
terminal, abbreviated as $d$ (see Appendix~\ref{app:dtd} for $d$'s
attributes).  Generated code checks character data conformance to the
type constraints.  This definition of $d$ is appropriate since BSML
does not support selections based on the type of character data.

One root non-terminal is initially generated for each schema block
(element, sequence, selection, repetition), each reference to a
primitive type, and each string of user code.  We denote non-terminals
by capital letters, the start non-terminal by $S$, the empty string by
$\epsilon$, and the root non-terminals generated for the children of
each schema block by $X_1,X_2,\ldots,X_n, n\ge 0$.  Further, lower-case
Greek letters denote (possibly empty) sequences of terminals,
non-terminals, and, in the next section, user codes.  With this
notation in mind, the definition of BSML is in
Figure~\ref{fig:bsml-def} (more details follow in future sections).  We
slightly deviate from a context-free grammar to allow for the
constraints on the number of repetitions (see next section).  To
reiterate, a grammar generated from a schema according to this
definition will undergo several standard equivalence transformations
before a grammar of the form shown in Figure~\ref{fig:octree-grammar}
is obtained.

\begin{figure}[|tbp|]
\begin{center}
\fbox{
$\begin{array}{l l c l}
\textrm{element}(id,opt,name,B_1,B_2,\ldots,B_n) &E& \rightarrow & s(name), X_1, X_2, \ldots, X_n, e(name) \\
&E&\rightarrow&\epsilon\quad\textrm{if\ } opt \\
\textrm{sequence}(id,opt,B_1,B_2,\ldots,B_n) &Q& \rightarrow & X_1, X_2, \ldots, X_n \\
&Q&\rightarrow&\epsilon\quad\textrm{if\ } opt \\
\textrm{selection}(id,opt,B_1,B_2,\ldots,B_n) &L& \rightarrow & X_1 \\
&L& \rightarrow & X_2 \\
& & & \cdots \\
&L& \rightarrow & X_n \\
&L&\rightarrow&\epsilon\quad\textrm{if\ } opt \\
\textrm{repetition}(id,opt,min,max,B_1,B_2,\ldots,B_n) &R& \rightarrow & \{B\}, X_1, X_2, \ldots, X_n, \{A\}, R', \{E\} \\
&R'& \rightarrow & X_1, X_2, \ldots, X_n, \{A\}, R' \\
&R'& \rightarrow & \epsilon \\
&R&\rightarrow&\epsilon\quad\textrm{if\ } opt \textrm{\ or\ } min=0 \\
\textrm{data}(base,min,max,number,finite,units) &D& \rightarrow & d(base,min,max,number,finite,units) \\
\textrm{code}(c) &C& \rightarrow & \{c\} \\
\end{array}$}
\end{center}
\caption{L-attributed definition of BSML.  Schema primitives, in a non-XML
notation, are on the left (see Figure~\ref{fig:octree-schema-nxml} for an
example) and their translations to grammar productions are on the right.
$B_1,B_2,\ldots,B_n$ are the children of the schema block and
$X_1,X_2,\ldots,X_n$ are the root non-terminals generated for
$B_1,B_2,\ldots,B_n$, respectively.  $opt$ is a boolean block
attribute; true means that the block is optional.  \{B\}, \{A\}, \{E\},
and \{c\} are binding codes explained in the next section.  References to
schema blocks (denoted by ref($id$)) are replaced with root non-terminals
of the blocks being referenced.  Definitions related to XML attributes are
omitted.}
\label{fig:bsml-def}
\end{figure}

The purpose of miscellaneous cleanup is to reduce the number of
non-terminals in the grammar.  These ad-hoc rewritings do not guarantee
that the resultant grammar is minimal in any strict sense.  Instead, they
address some inefficiencies that other steps are likely to introduce.
These cleanup steps were also chosen such that if the grammar were
LL(1) before cleanup, it would remain LL(1) after cleanup.  The
grammars shown in this paper have undergone two cleanup rewritings.
Each rewriting is applied until no further rewriting is possible.

\begin{enumerate}
\item Maximum length common suffixes are factored out.  $\beta\ne\epsilon$
is the maximum length common suffix of a non-terminal $A\ne S$ if (a) all
of $A$'s productions have the form $A\rightarrow\alpha_i\beta$,
$1\le i\le n$, (b) $\beta$ is of maximum length, and (c) neither
$\beta$ nor any $\alpha_i$ contain $A$.  If $n=1$, $A$ is eliminated
from the grammar and all occurrences of $A$ in the grammar are replaced
with $\beta$ ($\alpha_1=\epsilon$ because $\beta$ is of maximum length).
We call such non-terminals trivial.  Trivial non-terminals are often
introduced by schema-to-grammar conversion rules.  If $n>1$, all
occurrences of $A$ on the right-hand sides of all grammar productions
are replaced with $A\beta$ and the suffix $\beta$ is deleted from all
of $A$'s productions.  The purpose of this rewriting is to uncover
duplicate non-terminals for the next step.
\item Only one of any two duplicate non-terminals is retained.  Two
non-terminals $A\ne B$ are duplicate if whenever $A\rightarrow\alpha$
is in the grammar, $B\rightarrow\alpha$ is also in the grammar, and
vice versa.  $A$ is eliminated if $A\ne S$, $B$ is eliminated otherwise.
This definition is weak, e.g., $A$ and $B$ are not considered duplicate
if $A\rightarrow\alpha A\beta$ and $B\rightarrow\alpha B\beta$ are in the
grammar.  However, it suffices for our purposes.
\end{enumerate}

The expressive power of LL(1) grammars is well known.  In practice, the
limiting factor is not that the grammar is LL(1), but that the grammar
is annotated with user codes.  The next section gives two examples of
grammars that are not convertible to LL(1) because binding codes are
present.  A more interesting question is how the expressive power of
LL(1) grammars compares to the expressive power of BSML.  It is easy to
see that BSML can express a proper subset of LL(1) grammars.  For
example, $S\rightarrow s(x),e(y)$ is a valid LL(1) grammar, but BSML
cannot express it since no XML document that conforms to this grammar
is well-formed.

\paragraph{Observation 1.} Consider a subset of BSML that excludes
repetitions and user codes.  We say that BSML can express a grammar $G$
if a predictive parser generated from some schema in this restricted
subset of BSML can recognize precisely the language $L(G)$.  Clearly,
BSML cannot express any grammar $G$ that is not LL(1) (by construction
of the predictive parser).  Further, BSML cannot express an LL(1)
grammar $G$ unless:

\begin{enumerate}
\item if $d_1$ and $d_2$ are data terminals in $G$, then
$\forall\alpha,\beta: S\nRightarrow^+\alpha,d_1,d_2,\beta$
(data is atomic),
\item if $d$ is a data terminal and $S\Rightarrow^+\alpha,d,\beta$
is a derivation in $G$, then \\$\forall x,\gamma:\Big(
[\beta\nRightarrow^*s(x),\gamma]$ and
$[(\beta\Rightarrow^*e(x),\gamma)$ implies 
$(\forall y,\theta: \alpha\nRightarrow^*\theta,e(y))]\Big)$
(no mixed contents), and
\item if $s(x)$ is a start of element terminal, $g$ is $\epsilon$ or a data
terminal, and $S\Rightarrow^+\alpha,s(x),\beta$ is a derivation in $G$, then
$\Big([\beta\nRightarrow^*g]$ and
$[(y\ne x)$ implies $(\forall\gamma: \beta\nRightarrow^*g,e(y),\gamma)]\Big)$;
similarly, if $e(y)$ is an end of element terminal and
$S\Rightarrow^+\alpha,e(x),\beta$ is a derivation in $G$, then
$\Big([\alpha\nRightarrow^*g]$ and
$[(x\ne y)$ implies $(\forall\theta: \alpha\nRightarrow^*\theta,s(x),g)]\Big)$
(proper nesting of elements).\eobs
\end{enumerate}

The first two restrictions are specific to BSML and easy to relax.  However,
the last restriction is inherent in any XML schema language.  A good
schema language cannot describe documents that are not well-formed.
These are the necessary conditions, but it is not clear whether or not
they are sufficient.  We define schemas in terms of the schema language,
not in terms of LL(1) grammars, so converting from grammars to schemas
is not considered in this paper.

This section provided an overview of BSML features and defined BSML in
terms of an `almost context-free' grammar.  We outlined automatic
generation of predictive parsers that validate XML documents.  Further,
we have shown that the descriptive power of BSML is strictly less than
that of an LL(1) grammar where the terminals are SAX events.  The next
section extends validation to perform binding.

\section{Binding} \label{sec:binding}

Binding is a way to integrate semistructured data with languages
that were not designed to handle it (requirement~1).  Binding can
take several forms, depending on the language.  For FORTRAN and C,
binding usually means assigning values to language variables and
calling user-defined code to process these values (procedural binding).
It can also mean writing the data out in a format understood by the
component (format conversion).  For Matlab and SQL, binding entails
generating a script that contains embedded data and processing this
script by an interpreter (code generation).  The last two kinds of
binding can be thought of as XSLT-like transformations.

We implement all three kinds of binding by L-attributed definitions.
The schema language is extended by allowing user code to be injected
in the schema.  Schema languages that provide binding are called
\emph{binding schema markup languages}.  This section describes 
bindings in BSML and gives an example of their use.  Further, we
show how arbitrary binding codes limit the set of schemas supported
by BSML.

Let $c$ denote an arbitrary string of code.  Matching $\{c\}$ means
executing code $c$ while consuming no input tokens.  No assumptions are
made about the nature of $c$.  In particular, $c$ can (and usually
does) produce side effects, so $A\rightarrow\{c_1\},\{c_2\}$ and
$A\rightarrow\{c_2\},\{c_1\}$ can yield different results.  A
\emph{syntax-directed definition} is a context-free grammar extended by
allowing $\{c_j\}$ on the right-hand sides of productions.  For a
syntax-directed definition to be useful in binding, $c_j$ must contain
references to parts of the document being parsed.  We denote such
references by \verb|%x|, where \verb|x| is the id or the name of some
element or attribute.  When \verb|x| refers to an attribute or an
element of some primitive type, \verb|%x| is a value of the
attribute or the data contents of the element.  The type of \verb|%x|
is determined by the corresponding primitive type.  When \verb|x|
refers to an element of a wildcard type, \verb|%x| is a DOM
tree constructed from all descendants of \verb|x|, including itself.
This feature can be used for XHTML~\cite{xhtml} documentation.  The set
of attributes (elements) that are available to code $c$ depends on the
placement of $c$ in the syntax-directed definition and the parsing
strategy.  A syntax-directed definition is \emph{L-attributed} if, for
any derivation $S\Rightarrow^+\alpha\{c\}\beta$, any \verb|x|
referenced in $c$ is defined in all derivations of $\alpha$.  That is,
all attributes (elements) must be defined in a left-to-right scan
before they are referenced.  L-attributed definitions are easy to
implement with an LL(1) parser, but they restrict the set of grammars
reducible to LL(1).  Luckily, these restrictions are not important in
practice.

Figure~\ref{fig:pdp-matlab} gives an example binding schema for a PDP
(see Section~\ref{sec:example}) and Figure~\ref{fig:pdp-data} shows how
a parser generated from this schema converts a PDP encoded in XML to a
Matlab script.  This script will then be executed by an execution
manager (see Section~\ref{sec:integration}).  The same schema, with
different binding code, can convert an XML file to a number of SQL
INSERT statements that record the data in a relational database.  The
semantics of user codes are not limited to printing, so a FORTRAN
version of this binding can store the PDP in an array to be processed
later.  In other words, BSML bindings are compatible with any execution
environment that processes streams of data (requirement~7).  We use the
same approach to convert semistructured data to relational data, Matlab
scripts, and C structures.

The $\{B\}$, $\{A\}$, and $\{E\}$ codes in Figure~\ref{fig:pdp-matlab}
are generated for repetitions.  They are not necessary for this
example, but are required to enforce that each triangle has three
vertices in the previous example.  $\{B\}$ (begin repetition)
initializes the repetition count to zero.  Each repetition has its own
stack of counts.  $\{A\}$ (append) ensures that the maximum allowed
number of repetitions is not exceeded.  $\{E\}$ (end) checks the
minimum number of repetitions.  Thus, even simple validation (without
binding) is implemented in terms of an L-attributed definition, not
just an LL(1) grammar.

\begin{figure}[tbp]
\begin{center}

{\small\begin{verbatim}
<element name='pdp'>
  <element name='rds' optional='true' type='time' units='ns'/>
  <element name='med' optional='true' type='time' units='ns'/>
  <element name='pp' optional='true' type='power' units='dBW'/>
  <code>M=[</code>
  <repetition>
    <element name='ray'>
      <element name='time' type='time' units='ns'/>
      <element name='power' type='power' units='dBW'/>
    </element>
    <code>%time %power</code>
  </repetition>
  <code>];</code>
</element>
\end{verbatim}}

{\small $$\begin{array}{l l c l}
(S_1) & S & \rightarrow & s(pdp), R, M, P, \{\verb|M=[|\}, C, \{\verb|];|\}, e(pdp) \\
(R_1) & R & \rightarrow & \epsilon \\
(R_2) & R & \rightarrow & s(rds), d, e(rds) \\
(M_1) & M & \rightarrow & \epsilon \\
(M_2) & M & \rightarrow & s(med), d, e(med) \\
(P_1) & P & \rightarrow & \epsilon \\
(P_2) & P & \rightarrow & s(pp), d, e(pp) \\
(C_1) & C & \rightarrow & \epsilon \\
(C_2) & C & \rightarrow & \{B\}, s(ray), s(time), d, e(time), \\
      &   &             & s(power), d, e(power), e(ray), \\
      &   &             & \{\verb|%time %power|\}, \{A\}, X, \{E\} \\
(X_1) & X & \rightarrow & \epsilon \\
(X_2) & X & \rightarrow & s(ray), s(time), d, e(time), \\
      &   &             & s(power), d, e(power), e(ray), \\
      &   &             & \{\verb|%time %power|\}, \{A\}, X \\
\end{array}
\begin{array}{l|c c c c c c}
  & s(pdp) & s(rds) & s(med) & s(pp) & s(ray) & e(pdp) \\
\hline
S & S_1    &        &        &        &       &        \\
R &        & R_2    & R_1    & R_1    & R_1   & R_1  \\
M &        &        & M_2    & M_1    & M_1   & M_1  \\
P &        &        &        & P_2    & P_1   & P_1  \\
C &        &        &        &        & C_2   & C_1  \\
X &        &        &        &        & X_2   & X_1  \\
\hline
\end{array}$$}
\end{center}
\caption{(top) Binding schema for a power delay profile. $rds$, $med$,
and $pp$ stand for various optional statistics: rms delay spread, mean
excess delay, and peak power.  These statistics are ignored in this
example.  (left) L-attributed definition for a
power delay profile. $\{B\}$, $\{A\}$, and $\{E\}$ stand for codes
generated by the parser generator to handle repetitions.  Otherwise,
the meaning of $\{c\}$ is to print string $c$, followed by a new line
character, after expanding element references.  For clarity, full suffix
factoring was not performed, but trivial productions were eliminated.
(right) Predictive parsing table for a power delay profile.}
\label{fig:pdp-matlab}
\end{figure}

\begin{figure}[tbp]
{\small
\begin{minipage}{0.65\textwidth}
\begin{verbatim}
<pdp>
 <rds>23.0998</rds>
 <med>20.5691</med>
 <pp>-75.5665</pp>
 <ray><time>-4</time><power>-88.0937</power></ray>
 <ray><time>-3</time><power>-82.4416</power></ray>
 <ray><time>-2</time><power>-78.5346</power></ray>
 <ray><time>-1</time><power>-76.2634</power></ray>
 <ray><time>0</time><power>-75.5665</power></ray>
 <ray><time>1</time><power>-76.4908</power></ray>
 <ray><time>2</time><power>-79.2101</power></ray>
 <ray><time>3</time><power>-84.0673</power></ray>
 <ray><time>24</time><power>-86.4976</power></ray>
 <ray><time>25</time><power>-84.3451</power></ray>
 <ray><time>26</time><power>-84.3173</power></ray>
 <ray><time>27</time><power>-85.963</power></ray>
 <ray><time>28</time><power>-87.7374</power></ray>
 <ray><time>29</time><power>-88.6525</power></ray>
 <ray><time>43</time><power>-89.2007</power></ray>
 <ray><time>44</time><power>-83.17</power></ray>
 <ray><time>45</time><power>-79.2179</power></ray>
 <ray><time>46</time><power>-77.3306</power></ray>
 <ray><time>47</time><power>-77.4917</power></ray>
 <ray><time>48</time><power>-79.645</power></ray>
 <ray><time>49</time><power>-83.6205</power></ray>
 <ray><time>50</time><power>-88.7676</power></ray>
</pdp>
\end{verbatim}
\end{minipage}
\begin{minipage}{0.3\textwidth}
\begin{verbatim}



M=[
-4 -88.0937
-3 -82.4416
-2 -78.5346
-1 -76.2634
0 -75.5665
1 -76.4908
2 -79.2101
3 -84.0673
24 -86.4976
25 -84.3451
26 -84.3173
27 -85.963
28 -87.7374
29 -88.6525
43 -89.2007
44 -83.17
45 -79.2179
46 -77.3306
47 -77.4917
48 -79.645
49 -83.6205
50 -88.7676
];

\end{verbatim}
\end{minipage}
}
\caption{(left) An example PDP in XML.  The data corresponds to a
simulated channel in the corridor of the fourth floor of Durham Hall,
Virginia Tech.  The post processor samples the channel at 1~ns time
intervals to match the output of a channel sounder.  (right) Matlab
encoding of the PDP on the left, output by the parser generated from
the schema in Figure~\ref{fig:pdp-matlab}.} \label{fig:pdp-data}
\end{figure}

Unfortunately, L-attributed definitions make predictive parsing of
certain grammars impossible.  User codes can prevent elimination
of left recursion or left factoring of an L-attributed definition.
In the two examples below, grammars induced from the left-attributed
definitions by removing all user code can be transformed to LL(1).
However, the original L-attributed definitions cannot be transformed
to LL(1) without losing the stream semantics of the parser.

\paragraph{Example 1.} Consider a left-recursive schema and the
corresponding left-recursive grammar (after eliminating trivial
non-terminals):

\begin{center}
\begin{minipage}{0.6\textwidth}
\small{\begin{verbatim}
<selection id='s'> <sequence>
  <!-- empty -->
</sequence> <sequence>
  <code>c</code> <ref id='s'/>
  <element name='x'> <code>b</code> </element>
</sequence> </selection>
\end{verbatim}}
\end{minipage}
\begin{minipage}{0.35\textwidth}
$$\begin{array}{l c l}
S & \rightarrow & \epsilon \\
S & \rightarrow & \{c\}, S, s(x), \{b\}, e(x) \\
\end{array}$$
\end{minipage}
\end{center}

This grammar permits a derivation of the form
$S\Rightarrow^+\{c\}^k,(s(x),\{b\},e(x))^k$, $k>0$.  However, code $b$
cannot be executed before $k$ is known since $k$ executions of code $c$
must precede the first execution of code $b$.  Therefore, no LL(1)
parser with stream semantics can parse documents that conform to this
schema.  On the other hand, removing $\{c\}$ from the L-attributed
definition yields a grammar that is easily converted to LL(1):

$$\begin{array}{l c l}
S & \rightarrow & \epsilon \\
S & \rightarrow & S, s(x), \{b\}, e(x) \\
\end{array},\quad
\begin{array}{l c l}
S & \rightarrow & \epsilon \\
S & \rightarrow & s(x), \{b\}, e(x), S \\
\end{array}$$
This example is easy to generalize.\eobs

\paragraph{Observation 2.} Consider a set of all productions
for a non-terminal $A$.  Since any sequence $\{c_1\}\{c_2\}$ can be
rewritten as $\{c\}$, where $c=c_1 c_2$, we can uniquely represent this
set by a single production
$$A\rightarrow\{c_1\}A\alpha_1|\{c_2\}A\alpha_2|\cdots|\{c_n\}A\alpha_n
|\beta_1|\beta_2|\cdots|\beta_m,$$
where no $\beta_j, 1\le j\le m$, has a prefix $\{d\}A$.  Immediate left
recursion can be eliminated from this production without delaying user
code execution if and only if
\begin{enumerate}
\item $c_1=c_2=\cdots=c_n=\epsilon$ (no user code to the left) or
\item $\Big([(\beta_j\Rightarrow^*\gamma\{d\}\theta, 1\le j\le m)$ or
$(\alpha_i\Rightarrow^*\gamma\{d\}\theta, 1\le i\le n)]$ implies
$(d=\epsilon)\Big)$ (no user code to the right) and
$(c_1=c_2=\cdots=c_n)$ (same user code to the left).
\end{enumerate}
In all other cases, execution of user code must be delayed until
the last $\alpha_i$ is matched.\eobs

Consider a derivation of $A$ that
is no longer left-recursive (i.e., does not have a prefix of $\{d\}A$).
All such derivations can be written as
$$A\Rightarrow^{+}\{c_{i_1}\},\{c_{i_2}\},\ldots,\{c_{i_k}\},\beta_j,
\alpha_{i_k},\ldots,\alpha_{i_2},\alpha_{i_1},$$
where $\beta_j, 1\le j\le m$, stops left recursion after (at least)
$k+1$ steps and $1\le i_1,i_2,\ldots,i_k\le n$ represent the
choices for $\alpha_i$ in the derivation.  Suppose
$\beta_j\Rightarrow^*\gamma\{d\}\theta$ or
$\alpha_i\Rightarrow^*\gamma\{d\}\theta$.  The sequence of codes
$c_{i_1},c_{i_2},\ldots,c_{i_k}$ must be executed before code $d$, but
the LL(1) parser will only determine this sequence after it has parsed
all of $\beta_j,\alpha_{i_k},\ldots,\alpha_{i_2},\alpha_{i_1}$.  Thus,
eliminating left recursion entails delaying user code execution in all
but the trivial cases mentioned above.

\paragraph{Example 2.}  Left factoring of L-attributed definitions
poses similar problems.  Consider the following schema and L-attributed
definition (a more realistic version of this example would have a
repetition in place of the \verb|x| element):

\begin{center}
\begin{minipage}{0.6\textwidth}
{\small\begin{verbatim}
<selection> <sequence>
  <code>c</code>
  <element name='x'/><element name='y'/>
</sequence> <sequence>
  <code>d</code>
  <element name='x'/><element name='z'/>
</sequence> </selection>
\end{verbatim}}
\end{minipage}
\begin{minipage}{0.35\textwidth}
$$\begin{array}{l c l}
S & \rightarrow & \{c\}, s(x), e(x), s(y), e(y) \\
S & \rightarrow & \{d\}, s(x), e(x), s(z), e(z) \\
\end{array}$$
\end{minipage}
\end{center}
The decision about whether to execute code $c$ or $d$ cannot be made until
$s(y)$ or $s(z)$ is processed.  However, removing user codes makes this
L-attributed definition easy to refactor.  Again, we can show a more
general condition.\eobs

\paragraph{Observation 3.} Consider a set of all productions
for a non-terminal $A$ written as
$$A\rightarrow \alpha_1\beta_1|\alpha_2\beta_2|\cdots|\alpha_n\beta_n|
\gamma_1|\gamma_2|\cdots|\gamma_m,$$
such that $\alpha_1'=\alpha_2'=\cdots=\alpha_n'=\alpha\ne\epsilon$
($\alpha'$ denotes $\alpha$ with all user code removed)
and $\alpha$ is not a prefix of any $\gamma_1',\gamma_2',\ldots,\gamma_m'$.
Let the length of $\alpha$ be maximum and the lengths of
$\alpha_i, 1\le i\le n$, be minimum subject to $n\ge 2$, in which case
this representation of $A$ is unique.  $A$ can be left-factored without
delaying execution of user code if and only if
\begin{enumerate}
\item no rewriting of $A$ in the above form exists (no two definitions
of $A$ share the same prefix, less user codes), or
\item $\alpha_1=\alpha_2=\cdots=\alpha_n$ (same codes to the left) and
$A\rightarrow\gamma_1|\gamma_2|\cdots|\gamma_m$ can be left-factored.\eobs
\end{enumerate}

To summarize, we implement bindings in terms of L-attributed
definitions from parsing theory.  These bindings work well in practice,
but, in theory, annotating a schema that can be rewritten in LL(1) form
can make it no longer rewritable in LL(1) form.  This difficulty is
inherent in L-attributed definitions.  We currently assume that the
user is responsible for resolving such conflicts.  In practice, schemas
for PSE data rarely require complicated grammars.  Repetitions take
care of most of the recursive schema definitions.  To make LL(1)
parsing possible, troublesome content can be simply enclosed in an
extra XML element, whose start and end tags disambiguate the
transitions of the LL(1) parser.

\section{Conversion} \label{sec:conversion}

Conversion is the cornerstone of a system's ability to handle changes
and interface mismatches.  Conversion in a PSE helps to retain
historical data and facilitates inclusion of new components.  We use
change detection principles from~\cite{chawathe97}, with a few
important differences.  First, our goal is not merely to detect
changes, but to make PSE components work despite the changes.  Second,
we detect changes in the schema, not in the data.  The PSE environment
must guarantee that the data is in the right format for the component.
The job of the component is to process any data instance that conforms
to the right format.  Last, change detection and conversion are local
to the extent possible.  Locality is a virtue not only because it
allows for stream processing, but also because it limits sporadic
conversions between unrelated entities.

Similarly to the two previous sections, this section starts with a
comprehensive example.  Then, we describe the core of the conversion
algorithm and outline its limitations.  Finally, we extend the initial
algorithm to handle content replacements: unit conversion and
user-defined conversion filters.  At this point, it should not come as
a surprise to the reader that most of the technical limitations of
conversion are due to binding codes, not to the nature of the schema
language.  Therefore, the tedious details of handling binding codes are
omitted.  The emphasis is on non-technical limitations.  What forms of
semantic conversions can be `syntactized' in a schema language?  When
does such `syntactization' back fire and produce undesired outcomes?

The functional statement of the conversion problem can be given as
follows. Given the actual schema $S_a$ and the required schema $S_r$,
replace binding codes in $S_a$ with binding codes in $S_r$ and
conversion codes to obtain the conversion schema $S_c$.  $S_c$ must
describe precisely the documents described by $S_a$, but perform the
same bindings as $S_r$.

\paragraph{Example 3.}
Figure~\ref{fig:dif-schemas} depicts two
slightly different schemas for antenna descriptions in \SFW.  The
schema at the bottom (actual schema) was our first attempt at defining
a data format for antenna descriptions.  This version supported only
one antenna type and exhibited several inadequate representation
choices.  E.g., polar coordinates should have been used instead of
Cartesian coordinates because antenna designers prefer to work in the
polar coordinate system.  Antenna gain was not considered in the first
version because its effect is the same as changing transmitter power.
However, this seemingly unnecessary parameter should have been included
because it results in a more direct correspondence of simulation input
to a physical system.

The schema at the top of Fig.~\ref{fig:dif-schemas} (required schema)
improves upon the actual schema in several ways.  It better adheres to
common practices and supports more antenna types.  However, this schema
is different from the actual schema, while compatibility with old data
needs to be retained (requirement~2).  Figure~\ref{fig:conv-example}
illustrates how addition of conversion and binding codes to the actual
schema solves the compatibility problem.  A parser generated from the
conversion schema in Figure~\ref{fig:conv-example} will recognize the
actual data and provide the required binding.\eobs

\bigskip

\begin{figure}
\small{\begin{verbatim}
<element name='antennas'>
  <repetition>
    <element name='antenna'>
      <element name='id' type='string' min='1'/>
      <element name='phi' type='angle'/>
      <element name='theta' type='angle'/>
      <element name='gain' type='ratio' units='dB' optional='true' default='0'/>
      <code>puts stdout "%id: %phi %theta %gain"</code>
      <selection>
        <element name='waveguide'>
          <element name='width' type='distance' units='mm'/>
          <element name='height' type='distance' units='mm'/>
          <code>puts stdout "waveguide: %width %height"</code>
        </element>
        <element name='pyramidal_horn'>
          <element name='width' type='distance' units='mm'/>
          <element name='rw' type='distance' units='mm'/>
          <element name='height' type='distance' units='mm'/>
          <element name='rh' type='distance' units='mm'/>
          <code>puts stdout "pyramidal horn: %width %rw %height %rh"</code>
        </element>
      </selection>
    </element>
  </repetition>
</element>

<element name='antennas'>
  <repetition>
    <element name='antenna'>
      <element name='id' type='string' min='1'/>
      <element name='description' type='*'/>
      <element name='x' type='coordinate'/>
      <element name='y' type='coordinate'/>
      <element name='z' type='coordinate'/>
      <element name='waveguide'>
        <element name='width' type='distance' units='in'/>
        <element name='height' type='distance' units='in'/>
      </element>
    </element>
  </repetition>
</element>
\end{verbatim}}
\caption{Two slightly different schemas for a collection of antennas.
The component requires the top schema, but the data conforms to the
bottom schema.  The bottom schema (a)~represents antenna orientation
in Cartesian coordinates, not polar coordinates, (b)~lacks antenna
gain, (c)~requires antenna descriptions, (d)~measures antenna
dimensions in inches, not millimeters, and (e)~covers only one
antenna type.  The schema at the bottom does not contain binding
codes because they are irrelevant for this example.  All binding
codes are in Tcl.}
\label{fig:dif-schemas}
\end{figure}

\begin{figure}
\small{\begin{verbatim}
<element name='antennas'>
  <repetition>
    <element name='antenna'>
      <element name='id' type='string' min='1'/>
      <element name='description' type='*'/>
      <element name='x' type='coordinate'/>
      <element name='y' type='coordinate'/>
      <element name='z' type='coordinate'/>
      <code>  <!-- convert coordinates from rectangular to polar -->
        set _r [expr sqrt(%x*%x+%y*%y+%z*%z)]
        set %phi [expr atan2(%y,%x)]
        set %theta [expr acos(%z/$_r)]
      </code>
      <code>  <!-- set default gain -->
        set %gain 0
      </code>
      <code>puts stdout "%id: %phi %theta %gain"</code>
      <element name='waveguide'>
        <element name='width' type='distance' units='mm'/>
        <code>  <!-- convert units from inches to millimeters -->
          set %width [expr 25.4*%width]
        </code>
        <element name='height' type='distance' units='mm'/>
        <code>  <!-- convert units from inches to millimeters -->
          set %height [expr 25.4*%height]
        </code>
        <code>puts stdout "waveguide: %width %height"</code>
      </element>
    </element>
  </repetition>
</element>
\end{verbatim}}
\caption{Actual schema from Figure~\ref{fig:dif-schemas} (bottom) after
inserting conversion and binding codes.  This schema describes the actual
documents, but provides the bindings of the required schema
(top of Figure~\ref{fig:dif-schemas}).  We use {\tt\_r} instead
of {\tt\%r} because the latter could interfere with another use of
the name {\tt r}.}
\label{fig:conv-example}
\end{figure}

\begin{figure}
\begin{center}
\fbox{
$\begin{array}{@{}l l@{}}
D_r: & \textrm{data}(base_a,min_a,max_a,number_a,finite_a,units_a)\succeq
\textrm{data}(base_r,min_r,max_r,number_r,finite_r,units_r) \\
& \textrm{if\ } base_a=base_r, min_a\ge min_r, max_a\le max_r,
number_r\Rightarrow number_a, finite_r\Rightarrow finite_a, \\
& units_a=units_r \\
 & \\
E: & \textrm{element}(id_a,opt_a,name_a,C_{a1},C_{a2},\ldots,C_{an})\succeq
\textrm{element}(id_r,opt_r,name_r,C_{r1},C_{r2},\ldots,C_{rm}) \\
& \textrm{if\ } name_a=name_r, opt_a\Rightarrow opt_r,
Q_a(C_{a1},C_{a2},\ldots,C_{an})\succeq Q_r(C_{r1},C_{r2},\ldots,C_{rm}) \\
& \\
E_g: & X_a(id_a,opt_a,\ldots)\succeq \textrm{element}(id_r,opt_r,name_r,C_{r1},C_{r2},\ldots,C_{rm}) \\
& \textrm{if\ } opt_a\Rightarrow opt_r, Q_a(X_a(id_a,opt_a,\ldots))\succeq Q_r(C_{r1},C_{r2},\ldots,C_{rm}) \\
& \\
E_r: & \textrm{element}(id_a,opt_a,name_a,C_{a1},C_{a2},\ldots,C_{an})\succeq X_r(id_r,opt_r,\ldots) \\
& \textrm{if\ } opt_a\Rightarrow opt_r, Q_a(C_{a1},C_{a2},\ldots,C_{an})\succeq X_r(id_r,opt_r,\ldots) \\
& \\
P: & \textrm{sequence}(id_a,opt_a,C_{a1},C_{a2},\ldots,C_{an})\succeq
\textrm{sequence}(id_r,opt_r,C_{r1},C_{r2},\ldots,C_{rm}) \\
& \textrm{if\ } opt_a\Rightarrow opt_r,
Q_a(C_{a1},C_{a2},\ldots,C_{an})\succeq Q_r(C_{r1},C_{r2},\ldots,C_{rm}) \\
& \\
P_g: & X_a(id_a,opt_a,\ldots)\succeq\textrm{sequence}(id_r,opt_r,C_{r1},C_{r2},\ldots,C_{rm}) \\
& \textrm{if\ } opt_a\Rightarrow opt_r, Q_a(X_a(id_a,opt_a,\ldots))\succeq Q_r(C_{r1},C_{r2},\ldots,C_{rm}) \\
& \\
P_r: & \textrm{sequence}(id_a,opt_a,C_{a1},C_{a2},\ldots,C_{an})\succeq X_r(id_r,opt_r,\ldots) \\
& \textrm{if\ } opt_a\Rightarrow opt_r, Q_a(C_{a1},C_{a2},\ldots,C_{an})\succeq X_r(id_r,opt_r,\ldots) \\
& \\
C: & \textrm{selection}(id_a,opt_a,C_{a1},C_{a2},\ldots,C_{an})\succeq
\textrm{selection}(id_r,opt_r,C_{r1},C_{r2},\ldots,C_{rm}) \\
& \textrm{if\ } opt_a\Rightarrow opt_r,
\forall C_{ai}:(\exists! C_{rj}: C_{ai}\succeq C_{rj}) \\
& \\
C_g: & X_a(id_a,opt_a,\ldots)\succeq\textrm{selection}(id_r,opt_r,C_{r1},C_{r2},\ldots,C_{rm}) \\
 & \textrm{if\ }
opt_a\Rightarrow opt_r, (\exists! C_{rj}: X_a(id_a,opt_a,\ldots)\succeq C_{rj}) \\
& \\
R: & \textrm{repetition}(id_a,opt_a,min_a,max_a,C_{a1},C_{a2},\ldots,C_{an})\succeq
\textrm{repetition}(id_r,opt_r,min_r,max_r,C_{r1},C_{r2},\ldots,C_{rm}) \\
& \textrm{if\ } min_a\ge min_r, max_a\le max_r, opt_a\Rightarrow opt_r,
Q_a(C_{a1},C_{a2},\ldots,C_{an})\succeq Q_r(C_{r1},C_{r2},\ldots,C_{rm}) \\
& \\
R_g: & X_a(id_a,opt_a,\ldots)\succeq\textrm{repetition}(id_r,opt_r,min_r,max_r,C_{r1},C_{r2},\ldots,C_{rm}) \\
& \textrm{if\ } min_r\le 1, max_r\ge 1, opt_a\Rightarrow opt_r,
Q_a(X_a(id_a,opt_a,\ldots))\succeq Q_r(C_{r1},C_{r2},\ldots,C_{rm}) \\
& \\
F: & \textrm{ref}(id_a)\succeq\textrm{ref}(id_r) \\
& \textrm{if\ } X_a(id_a,opt_a,\ldots)\succeq X_r(id_r,opt_r,\ldots) \\
& \\
Q: & Q_a(C_{a1},C_{a2},\ldots,C_{an})\succeq Q_r(C_{r1},C_{r2},\ldots,C_{rm}) \\
& \textrm{if\ }
\forall C_{rj}(\ldots,opt_{rj},\ldots):[(\exists! C_{ai}: C_{ai}\succeq C_{rj}) \textrm{\ or\ } (opt_{rj})] \\
\end{array}$}
\end{center}
\caption{Version 1 of the `determines' relation
$X_a(id_a,opt_a,\ldots)\succeq X_r(id_r,opt_r,\ldots)$ between an actual
schema block $X_a(id_a,opt_a,\ldots)$ and a required schema block
$X_r(id_r,opt_r,\ldots)$.  We use the non-XML notation from
Figure~\ref{fig:octree-schema-nxml} plus $X_a(id_a,opt_a,\ldots)$ and
$X_r(id_r,opt_r,\ldots)$ are shortcuts for any schema block (data blocks
are never optional and have empty ids).  $\Rightarrow$ means logical
implication and $\exists!$ means `there exists a unique.'  The rules are
applied top to bottom, left to right.  The first matching rule wins (no
backtracking). This definition will be later restricted to make it
computable and rule~$Q$ will be extended to handle replacements.}
\label{fig:determines}
\end{figure}

\noindent
Following~\cite{chawathe97}, the basic assumption of the conversion
algorithm is that the actual schema $S_a$ can be converted to the
required schema $S_r$ by some sequence of `standard' edits. This
sequence of edits is called an \emph{edit script}.  Once the possible
types of edits are defined (what we can call a `conversion library'),
the job of the conversion algorithm is to (a)~find an edit script that
transforms the actual schema $S_a$ to the required schema $S_r$ and
(b)~express this edit script as data transformations, not schema
transformations.  In other words, the conversion algorithm looks for a
systematic procedure that converts actual data instances that conform
to $S_a$ to the required format $S_r$.  This procedure is expressed as
a conversion schema $S_c$ that has the structure of $S_a$, but binding
codes from $S_r$ and the conversion library.  $S_c$ is then used to
generate a parser that parses data instances conforming to $S_a$ and
acts as if it parsed data instances conforming to $S_r$.

Our conversion algorithm supports four kinds of schema edits:
\begin{enumerate}
\baselineskip=9.9pt
\itemsep=0.02in
\parsep=-0.5in
\item generalization,
\item restriction,
\item reordering, and 
\item replacement.
\end{enumerate}
\noindent
We use these terms in reference to the required schema, e.g., `the
required schema is a generalization of the actual schema.'
Generalization and restriction of schema trees are similar to
insertions and deletions in sequence alignment problems.  Reordering
and replacement mostly retain their standard meaning, except we
consider replacements of sets of schema blocks, not individual schema
blocks.  We first reduce the problem of converting trees to an easier
problem of converting sequences (see Figure~\ref{fig:determines}).
Sequence conversion (rule~$Q$) in this initial formulation performs all
conversions but replacements.  Then, we slightly restrict this
definition to make it practical and generalize rule~$Q$ to accommodate
replacements (unit conversion and user-defined conversion filters).

The conversion algorithm revolves around the `determines' relation
between schemas.  Intuitively, an actual schema $S_a$ should determine
a required schema $S_r$ if any document that conforms to $S_a$ contains
sufficient information to construct an `appropriate' document that
conforms to $S_r$.  `Appropriate' here is obviously a domain-specific
notion, and in the absence of a domain theory, there is no hard and
fast measure of `appropriateness.'  Given two slightly different
schemas, only a domain expert can tell whether or not it is meaningful
to attempt a conversion from one form to another. Therefore, our
conversion rules should be viewed as heuristics that we have found to
be useful enough to be supported in a conversion library. They are
neither sound nor complete in an algorithmic sense (because we do not
have an objective, external, measure of `conversion correctness').
Instead, they represent a tradeoff between soundness and completeness
and should be carefully evaluated for use in a particular domain. With
this disclaimer in mind, version~1 of the determines relation between
$S_a$ and $S_r$ ($S_a$ \emph{determines} $S_r$; $S_a\succeq S_r$) is
defined in Figure~\ref{fig:determines}.  We will also find the notion
of schema equivalence useful: we say that two schemas $S_a$ and $S_r$
are \emph{equivalent} if $S_a\succeq S_r$ and $S_r\succeq S_a$.

The first rule ($D_r$) in Figure~\ref{fig:determines}, for instance,
says that a value of primitive type (`data') can be substituted for
another if they have the same base type, their ranges are compatible,
and they have the same units.  It ensures that all primitive type
constraints of $S_r$ are met by $S_a$ (restriction).  Thus, $D_r$ is
simply a definition of type derivation by range restriction (the `r'
subscript in this and other rules stands for restriction; similarly,
the `g' subscript stands for generalization).  Rules $E$, $P$, and~$R$
state the obvious: two black boxes are compatible if they have
compatible wrappers (restriction) and compatible contents (any
conversions).  Rule~$C$ says that any choice in $S_a$ must uniquely
determine some choice in $S_r$ (restriction).  Rule~$Q$ enforces that
every block in $S_r$ is uniquely determined by some block in $S_a$.
This formulation of rule~$Q$ ignores extra blocks in $S_a$
(restriction), permits optional elements in $S_r$ to be unmatched
(generalization), and allows for contents reordering.  Rule~$F$ deals
with references.  Only rules $D_r$, $E$, $P$, $C$, and~$R$ are sound.
Rule~$F$ looks sound, but it makes the determines relation not
computable.  Rule~$Q$ is unsound primarily because it ignores
`unnecessary' blocks in $S_a$.

Rules $E_g$, $P_g$, $C_g$, and~$R_g$ handle generalizations across
schema blocks of (possibly) different types.  Their counterparts $E_r$
and~$P_r$ handle symmetric restrictions (why is there no $C_r$
or~$R_r$?).  Rule~$C_g$ was demonstrated in the example above.  It is a
base case for rule~$C$.  Rule~$C_g$ states that one way to generalize a
schema block is to enclose it in a selection, i.e., provide more
choices in $S_r$ than were available in $S_a$.  This rule is sound.
Rules~$E_g$, $P_g$, and~$R_g$ have similar motivations, but they are
unsound.  Essentially, we assume that decorating any black box with any
number of wrappers does not change the meaning of the black box
(generalization).  Similarly, we assume that wrappers can be freely
removed to expose the black box (restriction).

Consider a sequence of schemas that describes some physical system in
progressively greater detail.  Suppose some subsystem is described by
a single parameter.  Common practice is to allocate a single schema
block to this subsystem.  What happens when a more detailed description
of this subsystem is incorporated into the schema?  Chances are, the
original schema block allocated to the subsystem will be either
(a)~augmented with more contents (restriction part of rule~$Q$) or
(b)~wrapped in another block.  The generalization and restriction rules
handle case~(b).  However, blind application of these rules can lead to
disaster because these rules disregard some semantic information.
Examples will make these points clearer.

\paragraph{Example 4.} One common trick used to improve wireless system
performance is space-time transmit diversity (STTD).  Instead of a
single transmitter antenna, the base station uses two transmitter
antennas separated by a small distance.  PDPs are very sensitive to
device positioning, so two uncorrelated transmitter antennas can
produce widely different signals at the same receiver location.  If the
signal from one of the antennas is weak, the signal from another
antenna will probably be strong, so the overall performance is expected
to improve.  Consider how addition of STTD to the ray tracer affects
the schema of the transmitter file.  The original schema is on the left
and the new schema (with STTD support) is on the right.  The second
antenna is optional because STTD is not used in every system due to
cost considerations. 

\medskip

\noindent
(continued on next page)

\begin{center}
\begin{minipage}[b]{0.48\textwidth}
{\small\begin{verbatim}

<element name='tx'>
  <ref id='coordinates'/>
  <element name='power' type='power'/>
  <element name='freq' type='double'/>
</element>
\end{verbatim}}
\end{minipage}
\begin{minipage}{0.5\textwidth}
{\small\begin{verbatim}
<element name='base_station'>
  <element name='tx'>
    <ref id='coordinates'/>
    <element name='power' type='power'/>
    <element name='freq' type='double'/>
  </element>
  <element name='tx' optional='true'>
    <ref id='coordinates'/>
    <element name='power' type='power'/>
    <element name='freq' type='double'/>
  </element>
</element>
\end{verbatim}}
\end{minipage}
\end{center}

The new ray tracer should be able to work with old data because it
supports one or two transmitter antennas.  The old ray tracer should be
able to work with new data, albeit the results will be approximate when
the new data contains two transmitter antennas.  Further generalizing
this example to $n$ transmitter antennas would require a repetition.
We support conversion to repetitions, but not from repetitions.  For
this example, we could extract any antenna because they usually have
the same parameters and are positioned close together.  However, we
cannot extract an arbitrary ray from a PDP because the ray with maximum
power is usually intended.  Extracting any other ray would typically
produce nonsense results.\eobs

\paragraph{Example 5.} Havoc can result if rules $E_r$ and $E_g$ are
applied to the same element.  Element names have semantic meaning, but
this particular composition of rules allows arbitrary renaming of
elements.  Such renaming would make the following two schemas
equivalent.

\begin{center}
\verb|<element name='tx_gain' type='ratio'/>|$\quad$
\verb|<element name='snr' type='ratio'/>|
\end{center}

Even though both transmitter antenna gain and signal-to-noise ratio are
ratios measured in the same units (dB), they convey largely different
information.  We avoid such blatant mistakes by limiting the
application of generalization and restriction rules.  In particular, no
element can be renamed.\eobs

\bigskip

As the last example illustrates, the `determines' relation in
Figure~\ref{fig:determines} needs to be restricted.  It is helpful to
redefine this relation in terms of a context-free grammar that
describes $S_aS_r$.  Let the terminals be \verb|element(|,
\verb|sequence(|, \verb|selection(|, \verb|repetition(|, \verb|ref(|,
\verb|data(|, \verb|)|, and all element names and other values used in
two schemas under consideration.  Let the non-terminals be the labels
of the rules in Figure~\ref{fig:determines}, a special start
non-terminal $A$, and intermediate non-terminals introduced by the
rules.  We can formally define the necessary restrictions by limiting
the shape of the parse tree for $S_aS_r$.  Consider a path
$R_1,R_2,\ldots,R_n, n>0$, from some internal node $R_1\ne A$ to some
internal node $R_n\ne A$, where all $R_i, 1\le i\le n$, are rule
labels.  If $\mathcal{R}$ is the set of restriction rules and
$\mathcal{G}$ is the set of generalization rules, we require that
$(R_i\in\mathcal{R})$ implies $(R_{i-1}\notin\mathcal{G}$ and
$R_{i+1}\notin\mathcal{G})$, i.e., restriction and generalization rules
cannot be applied in sequence.  This restriction of the parse tree
disallows renaming of elements, but does not limit the number of
wrappers around black boxes.  Bounded determination deals with the
latter problem.  We say that $S_a$ $k$\emph{-determines} $S_r$
($S_a\succeq^kS_r$) if no path $R_1,R_2,\ldots,R_n$ contains a
substring of (possibly different) generalization (restriction) rules of
length greater than~$k$.  We leave it up to the reader to appropriately
restrict rule $F$ (reference).  These restrictions make the
`determines' relation computable and enforce locality of conversions.
As a side effect, we have shown that the problem of constructing a
conversion schema $S_c$ from the actual schema $S_a$ and the required
schema $S_r$ can be reduced to validation and binding (parsing and
translation).  However, schema conversion need not work with streams of
data, so a parser more powerful than a predictive parser should be
used.

It remains to consider requirements~4 and~5: unit conversion and
user-defined conversion filters (replacements).  Let~$D$ be a set of
all primitive types derived from double (recall that a primitive type
is defined by the base type, the range of legal values, and a unit
expression).  Unit conversion, e.g., converting kg/m$^2$ to lb/in$^2$,
is the simpler of the two replacements.  Both actual and required unit
expressions are converted to a canonical form (e.g., a fraction of
products of sums of CI units or dB) and then the conversion function is
found.  Unit conversions are functions of the form $$U:D_a\rightarrow
D_r,$$ where $D_a,D_r\in D$ are specific primitive types.  User-defined
conversion filters are functions of the form $$H:D_{a1}\times
D_{a2}\times \cdots\times D_{an}\rightarrow D_{r1}\times D_{r2}\times
\cdots\times D_{rm},$$ where $n,m>0$ and all $D_{ai},D_{rj}\in D,1\le
i\le n,1\le j\le m,$ are specific primitive types.  Arithmetic
operators and common mathematical functions are allowed in user-defined
conversion filters.  Each user-defined conversion filter is tagged with
element names $name_{a1},name_{a2},\ldots,name_{an}$ and
$name_{r1},name_{r2},\ldots,name_{rm}$ that determine when the filter
applies.  Such filters define rules of the form
$$\begin{array}{l}
(\textrm{element}(\$, \$, name_{a1}, D_{a1}),
\textrm{element}(\$, \$, name_{a2}, D_{a2}),
\ldots,
\textrm{element}(\$, \$, name_{an}, D_{an})) \succeq \\
(\textrm{element}(\$, \$, name_{r1}, D_{r1}),
\textrm{element}(\$, \$, name_{r2}, D_{r2}),
\ldots,
\textrm{element}(\$, \$, name_{rm}, D_{rm})). \\
\end{array}$$

Both kinds of filters are compiled into codes such as shown in
Figure~\ref{fig:conv-example}.  Rule~$Q$ is modified to take advantage
of replacements.  Basically, we are looking for (unique) partitions of
the actual schema blocks $C_{a1},C_{a2},\ldots,C_{an}$ and required
schema blocks $C_{r1},C_{r2},\ldots,C_{rm}$ such that each set of
schema blocks in the required partition is determined by some set of
schema blocks in the actual partition.  Determination can proceed
through the rules in Figure~\ref{fig:determines}, unit conversions,
and user-defined conversion filters (if everything else fails, optional
blocks in the required schema can remain unmatched).

%

The ultimate goal of the conversion algorithm is to find a meaningful
edit script.  However, this goal is impossible to achieve without
knowledge of the domain.  What happens when several edit scripts exist,
i.e., the problem of finding an edit script is ambiguous?  Depending on
the nature of the ambiguity, we can choose any edit script, the minimal
(in some sense) edit script, or to refuse to perform conversion.
The conversion algorithm described here either settles for some local
minimum (e.g., rule~$E$ is preferred over rule~$E_g$) or requires
uniqueness of conversions (rules $C$, $C_g$, and most of rule~$Q$).
Ambiguity remains an open problem that is unlikely to be solved by a
syntactic conversion algorithm.  Following the principle of least user
astonishment, we choose to reject most of ambiguous conversions.

Finally, let us consider how binding codes limit conversion.  We omit
formal treatment of the problem and limit the discussion to an
example.  It is easy to see that conversion may require delaying
binding code execution.  This should not be surprising since one kind
of conversion is reordering.

\paragraph{Example 6.}  Consider a required schema with binding codes
(left) and an actual schema (right).

\begin{center}
\begin{minipage}{0.45\textwidth}
{\small\begin{verbatim}
<sequence>
  <element name='a' type='double'/>
  <code>c1</code>
  <repetition>
    <ref id='b'/>
    <code>c2</code>
  </repetition>
<sequence>
\end{verbatim}}
\end{minipage}
\begin{minipage}{0.45\textwidth}
{\small\begin{verbatim}
<sequence>
  <repetition><ref id='b'/></repetition>
  <element name='x' type='double'/>
  <element name='y' type='double'/>
<sequence>
\end{verbatim}}
\end{minipage}
\end{center}

Assume that there exists a user-defined conversion filter that
calculates \verb|a| from \verb|x| and~\verb|y|.  If we ignore binding
code \verb|c2|, conversion is clearly local.  However, conversion with
\verb|c2| present will require delaying all executions of \verb|c2|
until \verb|c1| is executed.  The latter can only happen when the last
piece of the schema is matched.  In other words, binding codes should
be placed as late as possible in the schema.\eobs

This section presented a number of local conversions appropriate for
PSE data.  Conversions are carried out by extra codes injected in the
actual schema.  The conversion algorithm was built around the
`determines' relation between schemas.  The algorithm has some
technical limitations related to binding codes, but its major
limitation is conceptual.  Conversion, in the form presented here, is
syntactic.  It is based on the weak semistructured data model, not on
the underlying domain theory (wireless communications).  Therefore, we
can only speculate about the causes of differences between the actual
and required schemas.  There is no guarantee that automatic conversion
will produce meaningful results.  A stronger data model is necessary to
perform complex, yet meaningful, conversions.

\section{Integration with a PSE} \label{sec:integration}

\begin{figure}
\begin{center}
\includegraphics{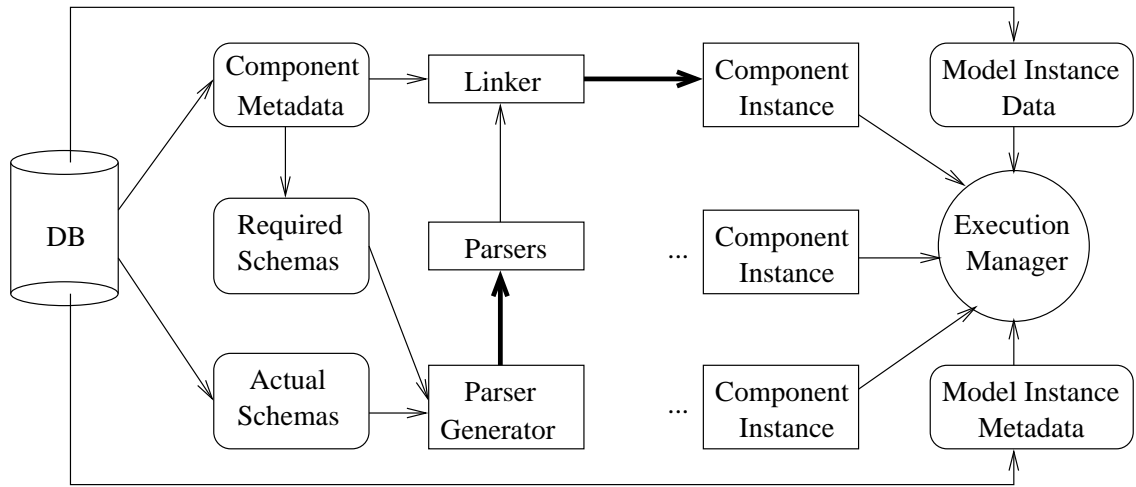}
\end{center}
\caption{BSML integration with PSE execution environment. The BSML parser
generator creates parsers that handle input ports of each component.
Execution manager controls the execution of a model instance that consists
of components, model instance data, and model instance metadata.
Figure~\ref{fig:loop1} partially defines one such instance.}
\label{fig:int}
\end{figure}

A complete PSE requires functionality far beyond validation, binding,
and conversion.  BSML ensures that the components can read streams of
XML data, but it does not support tasks such as scheduling,
communication, database storage and retrieval, connecting multiple
components into a given topology, and computational steering.  We
broadly call software that performs all of these tasks an
\emph{execution manager}. Figure~\ref{fig:int} illustrates how BSML
software and the execution manager function together.

From a systems point of view, BSML schemas are metadata and the BSML
software is a parser generator. Recall that the parser generator
generates parsers that perform validation, binding, and conversion functions
(every such generated parser will be able to take input data and stream it through
the component). Both the data and the metadata are
stored in a database.  We can distinguish three kinds of metadata:
schemas, component metadata, and model instance metadata.  Only one
form of metadata (schemas) was described in this paper.  Component
metadata contains component's local parameters, such as executable
name, programming language, and input/output port schemas.  It is the
kind of metadata used in CCAT.  Model instance metadata, i.e.,
component topology and other global execution parameters, serves a
purpose similar to GALE's workflow specifications.  It supports our
requirement~3.  

A parser is lazily generated for 
each used combination of component's
input port schema (required schema) and the schema of the data instance
connected to this port (actual schema).  Component metadata specifies
how linking must be performed (e.g., which of the three kinds of bindings
to use). Component instances are further managed by the execution manager.
Model instance metadata specifies how to execute the model 
instance (e.g., the topology and the number of processors), while 
model instance data serves as the actual (data) input to the model instance.  
To summarize, the BSML parser generator creates
component instances---programs that take a number of XML streams as
inputs and produce a number of XML streams as outputs.  This
representation is appropriate for management of a PSE execution
environment.

\subsection{Status of Prototype}
In \SFW, the execution manager is implemented in Tcl/Tk and most of the
component metadata is hard-coded.  Model instance metadata consists
primarily of the number of processors and a cross-product of references
to model instance data.  An (incomplete) example of such a
specification is 
\begin{quote}
`compute power coverage maps for these three
transmitter locations in Torgersen Hall and show a graph of BERs with
the signal-to-noise ratio varying from zero to twenty dB in steps of
two dB; use thirty nodes of a 200-node Beowulf cluster.' 
\end{quote}
PostgreSQL and
the filesystem serve the role of the database.  Large files (e.g.,
floor plans) are typically stored in the filesystem and small ones
(e.g., PDPs) are usually imported into PostgreSQL.  The parser
generator is written in SWI Prolog.  It generates parsers in Tcl.
Currently, these parsers are used mostly in the execution manager,
visualization components, and database interfacing components.

\section{Discussion}
\label{sec:discussion}
We have described the use of validation, binding, and conversion facilities to
solve data interchange problems in a PSE.  Since all three concepts are
closely related to parsing and translation, viewing application
composition in terms of data management uncovers well-understood
solutions to interface mismatch problems.  The semistructured data
model allows us to syntactically define several forms of conversions
that are usually implemented by hand-written mediators in PSEs.  Such
automation reduces the cost of PSE development and, more importantly,
brings PSEs closer to their ultimate goal --- namely, PSE users should be solving
their domain-specific problems, not be beset by the technical details of component
composition in a heterogeneous computing environment.

Several extensions to the present work are envisioned.  First,
the expressiveness of schema languages for data interchange and application
composition can be formally characterized. This will allow us to 
reason about requirements such as stream processing from a modeling
perspective. Such a study will also lead to a better understanding of the
roles that a markup language can play in a PSE.
Second, dataflow relationships between components can 
be made explicit.  BSML guarantees that any component
instance be able to process streams of data, but synchronization issues are meant
to be resolved by the execution manager. Tighter integration of
BSML and composition frameworks can be explored.
Finally, the overall view of a PSE as a semistructured data management system deserves further
exploration.  For example, it seems possible to automatically generate
workflow specifications from queries on a semistructured database of
simulation results.  

Any good problem solving facility is characterized by `what it lets you get away with.'
BSML is unique among PSE projects in that it allows a modeler or engineer to
flexibly incorporate application-specific considerations for data interchange,
without insisting on an implementation vocabulary for components.


\bibliographystyle{plain}
\bibliography{paper}

\appendix

\section{BSML DTD} \label{app:dtd}

{\small\begin{verbatim}
<!ENTITY % boolean "(true|false|t|f|yes|no|y|n)">

<!-- attributes of primitive types:
  min - minimum value or string length (inclusive)
  max - maximum value or string length (inclusive)
  number - true means NaN is not allowed (doubles only)
  finite - true means +/-infinity is not allowed (doubles only)
  units - units for this type (doubles only)
-->
<!ENTITY % type_attributes "
    min         CDATA       #IMPLIED
    max         CDATA       #IMPLIED
    number      %boolean;   #IMPLIED
    finite      %boolean;   #IMPLIED
    units       CDATA       #IMPLIED
">

<!-- what schemas and schema blocks are composed of -->
<!ENTITY % schema_contents "
    (element | sequence | selection | repetition)
">
<!ENTITY % block_contents "
    (%schema_contents; | default | ref | code)
">


<!-- a collection of schemas -->
<!ELEMENT schemas ((description)?, (type | schema)*)>
<!ATTLIST schemas>

<!-- primitive type: attributes above and an optional
enumeration of legal values; derivation works by restriction;
builtin base types are: integer, string, double, boolean -->
<!ELEMENT type ((description)?, (values)?)>
<!ATTLIST type
    id          CDATA       #REQUIRED
    base        CDATA       #REQUIRED
    %type_attributes;
>
<!-- enumeration of legal values, no value is legal if empty -->
<!ELEMENT values ((value)*)>
<!ATTLIST values>
<!ELEMENT value (#PCDATA)>
<!ATTLIST value>

<!-- schema -->
<!ELEMENT schema ((description)?, (code)*, (%schema_contents;), (code)*)>
<!ATTLIST schema
    id          CDATA       #REQUIRED
>

<!-- an element can contain either
 (a) character data of a primitive type (type attribute is present),
 (b) zero or more schema blocks (type attribute is absent), or
 (c) when type='*', any contents.
-->
<!ELEMENT element ((description)?, (attribute)*,
                   ((values)? | (%block_contents;)*))>
<!ATTLIST element
    name        CDATA       #REQUIRED
    id          CDATA       #IMPLIED
    optional    %boolean;   "false"
    type        CDATA       #IMPLIED
    %type_attributes;
    default     CDATA       #IMPLIED
>

<!-- an attribute must contain a value of some primitive type -->
<!ELEMENT attribute ((description)?, (values)?)>
<!ATTLIST attribute
    name        CDATA       #REQUIRED
    id          CDATA       #IMPLIED
    type        CDATA       "string"
    %type_attributes;
    default     CDATA       #IMPLIED
>


<!-- a sequence is just a grouping, for convenience -->
<!ELEMENT sequence ((description)?, (%block_contents;)*)>
<!ATTLIST sequence
    id          CDATA       #IMPLIED
    optional    %boolean;   "false"
>

<!-- a selection denotes a mutually exclusive choice of contents -->
<!ELEMENT selection ((description)?, (%block_contents;)+)>
<!ATTLIST selection
    id          CDATA       #IMPLIED
    optional    %boolean;   "false"
>

<!-- a repetition denotes [min..max] repetitions of contents -->
<!ELEMENT repetition ((description)?, (%block_contents)*)>
<!ATTLIST repetition
    id          CDATA       #IMPLIED
    optional    %boolean;   "false"
    min         CDATA       "0"
    max         CDATA       "inf"
>

<!-- a reference to some block id in this schema,
or to an id of a different schema -->
<!ELEMENT ref ((description)?)>
<!ATTLIST ref
    id          CDATA       #REQUIRED
>

<!-- user code; language and component attributes facilitate
schema reuse (different components can have the same schema,
but different binding codes) -->
<!ELEMENT code (#PCDATA)>
<!ATTLIST code
    language    CDATA       #IMPLIED
    component   CDATA       #IMPLIED
>

<!-- default contents must conform to BSML schema block -->
<!ELEMENT default ANY>

<!-- XHTML usually goes here -->
<!ELEMENT description ANY>
\end{verbatim}}

\end{document}